\documentclass{emulateapj}

\usepackage{natbib}
\usepackage{epsfig}
\usepackage{amssymb,amsmath}
\usepackage{color}
\usepackage{xspace}

\shorttitle{Automated Lens Detection}
\shortauthors{Marshall et~al.}

\def\ifm#1{\relax\ifmmode#1\else$\mathsurround=0pt #1$\fi}

\def\ltsima{$\; \buildrel < \over \sim \;$}
\def\lsim{\lower.5ex\hbox{\ltsima}}
\def\gtsima{$\; \buildrel > \over \sim \;$}
\def\gsim{\lower.5ex\hbox{\gtsima}}
\def\eg{{e.g.}\xspace}
\def\ie{{i.e.}\xspace}

\def\data{\mathbf{d}}
\def\pr{{\rm Pr}}
\def\thetaE{\theta_{\rm E}}
\def\truethetaE{\hat{\theta}_{\rm E}}
\def\gammaX{\gamma_{\rm X}}
\def\phiX{\phi_{\rm X}}
\def\dthetaE{\delta\theta_{\rm E}}
\def\nsigma{n_{\sigma}}
\def\Rd{R_{\rm d}}
\def\ms{m_{\rm s}}
\def\truems{\hat{m}_{\rm s}}
\def\dms{\delta m_{\rm s}}
\def\Hr{H_{\rm r}}

\def\hst{{\it HST}\xspace}

\begin{document}

\title{Automated detection of galaxy-scale gravitational lenses \\
       in high-resolution imaging data}

\author{Philip~J.~Marshall\altaffilmark{1,2}}
\author{David~W.~Hogg\altaffilmark{3}}
\author{Leonidas~A.~Moustakas\altaffilmark{4}}
\author{Christopher~D.~Fassnacht\altaffilmark{5}}
\author{Maru\v{s}a~Brada\v{c}\altaffilmark{1,2,*}}
\author{Tim~Schrabback\altaffilmark{6,7}}
\author{Roger~D.~Blandford\altaffilmark{2}}

\altaffiltext{1}{Physics department, University of California, Santa Barbara, CA 93601, {\tt e-mail: pjm@physics.ucsb.edu}} 
\altaffiltext{2}{KIPAC, P.O. Box 20450, MS29, Stanford, CA 94309} 
\altaffiltext{3}{Center for Cosmology and Particle Physics, Department of Physics, New York University, 4 Washington Place, New York, NY 10003}  
\altaffiltext{4}{JPL/Caltech, 4800 Oak Grove Dr, MS\,169-327, Pasadena, CA 91109}
\altaffiltext{5}{Department of Physics, U.C.~Davis, Davis, CA 95616} 
\altaffiltext{6}{Argelander-Institut f\"ur Astronomie, Universit\"at Bonn, Auf dem H\"ugel 71, 53121 Bonn, Germany} 
\altaffiltext{7}{Leiden Observatory, Leiden University, Niels Bohrweg 2, 2333 CA Leiden, The Netherlands} 
\altaffiltext{*}{Hubble Fellow}

%-------------------------------------------------------------------------------

\begin{abstract} 
Direct lens modeling is the key to successful and meaningful automated
strong galaxy-scale gravitational lens detection. We have implemented
a lens-modeling ``robot'' that treats every bright red galaxy (BRG) in
a large imaging survey as a potential gravitational lens system.
Using a simple model optimized for ``typical'' galaxy-scale
gravitational lenses, we maximize the multiply-imaged source plane
flux and generate, for the resulting best lens model, four assessments
of model quality that are then used in an automated
classification.  The robot infers from these four data the lens
classification parameter~$H$ that a human would have assigned; the
inference is performed using a probability distribution
generated from a human-classified training set of candidates,
including realistic simulated lenses and known false positives drawn
from the \hst Extended Groth Strip (EGS) survey.
We compute the expected purity, completeness and rejection rate, and
find that these statistics can be optimized for a particular
application by changing the prior probability distribution 
for $H$; this is equivalent to
defining the robot's ``character.''  Adopting a realistic prior based
on expectations for the abundance of lenses, we find that a lens
sample may be generated that is $\sim100$\% pure, but only $\sim20$\%
complete.  This shortfall is due primarily to the over-simplicity of
the lens model. With a more optimistic robot, $\sim90$\% 
completeness can
be achieved while rejecting $\sim90$\% of the candidate objects. The
remaining candidates must be classified by human inspectors.
Displaying the images used and produced by the robot on a custom
``one-click'' web interface, we are able to inspect and classify lens
candidates at a rate of a few seconds per system, suggesting that a
future 1000 square degree imaging survey containing $10^7$ BRGs, and
some $10^4$ lenses, could be successfully, and reproducibly, searched
in a modest amount of time.  We have verified our projected survey
statistics, albeit at low significance, using the \hst/EGS data,
discovering four new lens candidates in the process.

\end{abstract}


%-------------------------------------------------------------------------------

\section{Introduction}\label{sec:intro}

Large, well-defined samples of strong gravitational lenses enable many
important astrophysical and cosmological investigations.  These
include measuring the projected dark and luminous mass distributions
of galaxies \citep[\eg][]{T+K04,Koo++06}, measuring the expansion rate
of the universe $H(z)$ with lens time delays
\citep[\eg][]{Suy++08,Ogu07}, and also cosmological volumes and
distance ratios \citep[\eg][]{Mit++05,C+N07}, and probing the dark
galaxy substructure predicted by CDM models \citep{K+D04,Bra++04}.
Extended lensed source galaxy images provide more constraints on the
lens mass distribution \citep[\eg][]{Koo05} than do lensed point
sources.  Since lensing conserves surface brightness, the distant
lensed (``source'') objects are greatly extended and magnified,
revealing details that may not be otherwise visible. For example,
recent observations using gravitational lenses as ``cosmic
telescopes'' have provided unprecedented views of forming dwarf
galaxies \citep[\eg][]{Mar++07}, Lyman-break galaxies at 
$z=2-4$ \citep[see \eg][]{BMD00,Sma++07,All++07}, and quasar accretion
disks~\citep[\eg][]{PMK08}. Having a large array of cosmic telescopes
will allow us to select the best ones for any given application, but
will also allow us to build up a statistical picture of the source
population.

Since the first strong lens was discovered,
\citep[Q\,0957+561,][]{WCW79}, the $\sim$\,200 galaxy-scale lenses
known have been found by a combination of serendipitous
discovery and a host of systematic search techniques.  These
techniques include visual inspection of deep optical imaging obtained
with the {\em Hubble Space Telescope} \citep[{\hst};
  \eg][]{Hog++96,ZMD97,RGO99,Fas++04,Mou++07, Fau++08}, targeted
imaging of the population of potentially lensed quasars or radio
sources
\citep[\eg][]{Mye++03,Bro++03,Ina++03,Ogu++04,Pin++06}, and followup
of systems for which optical spectroscopy revealed anomalous emission
lines \citep[\eg][]{War++96,Bol++04,Wil++06}. Further techniques using
time-domain information have been proposed as efficient lens finders
\citep[\eg][]{Pin05,Koc++06}.

It is a spectroscopic survey that has generated the largest single
lens sample to date. 
With the SDSS galaxy redshift survey as its feeder, the
Sloan-Lens ACS Survey (SLACS) project \citep{Bol++06} has discovered
$\sim70$ new galaxy-scale strong lenses~\citep{Bol++08}.  Candidates
are selected for their background galaxy emission lines present in the
foreground old stellar population spectrum, and then confirmed using
high resolution \hst imaging. The SDSS selection function limits the
median lens redshift to be $\approx 0.2$, 
while the need for detectable emission
lines hides many lenses and incurs a strong 
``magnification bias''
towards ring-like systems. Future surveys will extend the lens
redshift limit somewhat, but likely not reach the same area as SDSS.

Instead, we anticipate that samples of galaxy-scale gravitational
lenses that are 2-3 orders of magnitude larger than the current set
will come, at least before SKA, from large optical imaging
surveys. For example, the proposed Super-Nova Acceleration Probe
(SNAP) mission is slated to include a 1000 square degree survey at
\hst/WFPC2 resolution \citep{Ald++04,MBS05}, with multi-band imaging
to provide photometric redshifts of faint galaxies, and discover some
$10^4$ strong lenses. From the ground, KIDS, Pan-STARRS, DES and LSST will
all make significant advances in strong lensing, but will be limited
by their angular resolution: their strengths will lie in the finding
of large numbers of time-varying, and/or wide-separation, lensed
systems. The majority of the strong lensing cross-section in the
universe lies in massive elliptical galaxies \citep{TOG84}.
Therefore, in a high-resolution space-based survey, the majority of
lenses will be massive elliptical galaxies with redshifts 0.5--1.0,
multiply-imaging faint blue star-forming galaxies at redshifts 1.0 and
above~\citep[\eg][]{MBS05}. 
This suggests that an efficient strategy is
to focus on bright red galaxies (BRGs) 
as being the ``typical'' potential
lenses~\citep[\eg][]{Fas++04,Fau++08}.

Cost limitations mean we may not be able to perform spectroscopy on
all future lens candidates, whose sources are likely to be quite
faint, and we will need to rely on a better, more quantitative
understanding of the imaging data in hand.  Indeed, with only image
information available, the classification of any lens candidate must
come entirely from modeling. Does a model for the image where some of
the features are lensed by a massive object (consistent with the
observed elliptical galaxy) explain the data? Are the residuals from
this modeling process consistent with what we know about early-type
galaxy structure? We suggest that the optimal way to \emph{find}
lenses in optical imaging surveys is to classify objects by their
ability to be modeled as gravitational lenses.

In the imaging survey gravitational lens searches already carried out,
the lens modeling has been performed \emph{after} a sample of
candidates has been generated by other means, and classified by
experienced human inspectors.  Indeed, this human inspection stage can
be sufficiently effective 
that it can be thought of as an approximate 
lens-modeling process.  
Present-day \hst surveys of area~$\sim$\,1~square
degree are small enough that human inspection is tedious but feasible.
However, looking forward to the 
next generation of high resolution 
imaging surveys, and desiring
to {\em build on} the extensive human expertise in identifying lenses,
we are motivated to develop an automated lens-finding ``software
robot.''  The tirelessness of this robot would enable it
to find lenses with a well-understood, calculable, and
reproducible selection function; this will be a vital property for the
resulting lens samples to be 
statistically useful.

In this work, we describe just such a robot: it models every possible
candidate massive galaxy image in the survey as a combination of
foreground (lensing) galaxy and multiply-imaged background (lensed)
source, and then interprets the results based on its previous
experience with both real and simulated data. We will argue that, at
least at
present, the most effective automated methods mimic the operation
of a human analyst: our robot attempts to predict, via a probabilistic
model, the classification that would have been made by a human.

We can prepare for a future of much larger surveys by working with
manageably-sized current samples of lens candidates from extant high
resolution imaging surveys. The \hst archive contains several square
degrees of suitably-surveyed (\eg\ deep, high galactic latitude) sky,
which we are searching for serendipitous lensing events \citep[the
  HAGGLeS project, program HST-AR-10678][Marshall et al.\ in
  preparation]{Mar++05}. As a pilot for this project, we carried out a
by-eye search for lenses in the 0.17 square degree area of the
\hst/ACS coverage of the Extended Groth Strip \citep[EGS,][]{Dav++07},
finding three definite lenses \citep[][hereafter M07]{Mou++07}, which
includes one that was found (also by eye) in the \hst Medium Deep
Survey \citep{RGO99}, and~four candidates of lower
believability~\citep[][hereafter M06]{Mou++06}. This dataset therefore
makes an excellent testing ground for new lens detection methods.

This paper is organized as follows. In
Sections~\ref{sect:method:modeling} (lens modeling)
and~\ref{sect:method:classification} (probabilistic interpretation) we
outline our lens-finding algorithm, and describe its implementation as
a software robot.  In Section~\ref{sect:training} we educate the
robot, using a large set of simulated galaxy-scale lenses, and a
roughly equal number of human-classified non-lenses from the \hst EGS
survey.  As a step towards understanding the selection function of
galaxy-galaxy strong lenses, we assess the completeness and purity of
the robot-generated sample at this point.  Then in
Section~\ref{sect:egs}, as a ``blind test'' we apply our automatic
lens finder to the remaining EGS images used in M06 and M07, and
compare with the by-eye search results. Following some discussion of
the strengths and weaknesses of our approach in
Section~\ref{sect:discuss}, we present our conclusions in
Section~\ref{sect:concl}. All magnitudes referred to are calculated in
the AB system.

%-------------------------------------------------------------------------------

\section{Lens modeling}\label{sect:method:modeling}

From a given potential lens galaxy, suitably selected (\eg\ by color
and magnitude) and extracted from the survey images, we subtract a
smooth, symmetric model of its own intensity, leaving a residual map
that may potentially (and perhaps partially) be explained by multiple
images of a background source.  For each possible lens model in a
finite, gridded space, we compute gravitational lens deflections that
are used to transform the observed (``image-plane'') pixels to their
nearest-pixel unlensed (``source-plane'') positions.  Because the
models under consideration are multiply-imaging, more than one
image-plane pixel may map to each source-plane pixel; since
gravitational lensing conserves surface brightness, the pixels of the
image plane mapping to the same point in the source plane should have
equal values. To optimize the model we construct a scalar that is
maximized when non-trivial fractions of image-plane pixels ``agree''
on the intensity to be assigned to these source-plane pixels.  The
bright and dark parts of the residual maps both participate in this
agreement; indeed, blank sky in the image plane where a second, third,
or fourth image ought to appear, given structure in the source plane, is
more damning to a possible lens model than image-plane flux where no
image is expected, and our scalar metric captures this. 

The method is motivated by the observation that when multiply-imaged
galaxies are faint,  the best evidence that they are indeed
multiply-imaged is that there is a reasonable gravitational lens model
with a reasonable unlensed intensity pattern that describes the
morphology as observed.  

In order to make best use of the reader's attention span we give a
very brief overview of our lens modeling procedure below, and then
expand on the technical details in
subsection~\ref{sect:method:modeling:technical}. We expect many might
discreetly push ahead to Section~\ref{sect:method:classification} at
this point.

% - - - - - - - - - - - - - - - - - - - - - - - - - - - - - - - - - - - - - -

\subsection{Overview of modeling procedure}\label{sect:method:modeling:overview}

\begin{itemize}

\item We assume that most strong lensing galaxies have the smooth,
  symmetric morphologies of early-type galaxies, and subtract off the
  best-fitting set of concentric elliptical isophotes. In practice we
  fit an elliptically-symmetric Moffat profile, which is very
  effective at removing galaxy bulges. Many bright red galaxies do
  show some disk component in their images; we identify these by their
  position and orientation relative to the major axis of the bulge,
  and mask out all pixels associated with the features.

\item We assume that these galaxies' lens potentials can be modeled
  adequately with singular isothermal sphere models plus external
  shear, and we grid this 3-dimensional parameter space with a coarse
  pixelization. The motivation for using this simple model is partly
  empirical, based on the results from the SLACS project
  \citep{Koo++06}, and partly to keep the CPU time manageable.

\item We assume that the background source sizes, or angular scale
  over which background sources vary, is comparable to, or larger than
  the \hst point spread function (PSF), and consequently ignore the
  latter when tracing flux back to the source plane. With any given
  lens model we perform a very crude ray-tracing to map image plane
  pixels onto a (finer) grid of source pixels.
  
\item In any putative multiple-imaging system, we require (for
  identification as a lens) that at least some of the detected
  residuals lying in the multiply-imaged part of the image plane be
  explained by a multiply-imaged background source. 
  
\item When mapping a set of multiply-imaged image-plane pixels back to
  the same source plane pixel, we take the minimum value of the image
  pixels' fluxes. This allows us to use all the information in the
  image, using dark sky to veto non-lensed bright sky.
  
\item Stepping through the entire grid of models, we select the one
  that maximizes this ``minimum-filtered'' source plane flux as our
  ``optimal'' model.  
  
\end{itemize}

% - - - - - - - - - - - - - - - - - - - - - - - - - - - - - - - - - - - - - -

\subsection{Technical notes}\label{sect:method:modeling:technical}

The image-plane and source-plane pixel grids we choose are determined
by the pixelization of the input image. We use drizzled \hst/ACS images
with pixel scale 0.03\,arcsec, and investigate 6\,arcsec cutout images
centered on a pre-selected elliptical galaxy position. This selection
is an important part of the lens search: one may consider using both
color and morphology to pre-select elliptical galaxies, although this
process is made considerably harder by the (possible) presence of the
contrasting lensed images themselves. In this work we adopt the
inclusive approach of investigating all bright extended objects, and
exploring the performance of our technique on the selection criteria
involved. To reduce computation time, and improve the accuracy of the
ray-tracing, we bin the image plane by a factor of two. This reduces
the number of deflection angles to be calculated by a factor of four
(and further justifies our ignoring the PSF at this stage).

The grid of model space we choose for the lensing potential is
three-dimensional: we use a singular isothermal sphere (SIS) with
external shear \citep[\eg][]{Koc91,KSB94}. The primary parameter is
the SIS lens strength, specified as the angular 
Einstein radius $\thetaE$. 
Having a more subtle effect on the lensing
behavior are the two parameters that describe the magnitude and
direction of the external shear, $\gammaX$ and $\phiX$. These two
parameters are vital, since a significant fraction of all known lenses
are four image systems (quads). Without the external shear the SIS
model can only produce two images.  All points in this parameter space
are treated as being equally likely prior to fitting each source, and
are then stepped through in an exhaustive search. The coarseness of
the grid was chosen to keep the CPU time per system low. We restricted
the Einstein radius to lie between 0.4 and 1.8\,arcsec, large enough
to include all SIS lenses with velocity dispersion in the range 160 to
350\,km/s (given a lens at $z_{\rm d}=0.5$ and a source at redshift
$z_{\rm s}=1.0$).

For each lens model we trace the image plane pixel values back to the
source plane via the usual lens equation,
\begin{equation}
\boldsymbol{\beta} = \boldsymbol{\theta} - \boldsymbol{\alpha}
\end{equation}
where the optical axis was taken to be the centroid of the elliptical
galaxy light, and the two components of the deflection angle
$\boldsymbol{\alpha}$ are given by
\begin{align}
\alpha_1 &= \theta_1\frac{\thetaE}{|\boldsymbol{\theta}|} + 
  \gammaX \left( \theta_1\cos{2\phiX} + \theta_2\sin{2\phiX} \right) \\
\alpha_2 &= \theta_2\frac{\thetaE}{|\boldsymbol{\theta}|} + 
  \gammaX \left( \theta_1\sin{2\phiX} - \theta_2\cos{2\phiX} \right)
\end{align}
We use nearest-neighbor mappings; that is, we ``snap'' the lensing
deflections ``to grid.'' This is a time-saving device: 
rather than having square pixels in the image
plane map to distorted rectangles in the source plane, we are crudely
approximating the source using a square grid. We will discuss the
effects of this approximation in Section~\ref{sect:training:calib};
here we note that we do at least use a source plane that is twice
as finely gridded as the (binned) image plane.

We are specifically interested in multiple-image systems:
some source-plane pixels will have more than one
associated image-plane pixel mapping to each one, 
and in general those
associated image-plane pixels will come from widely separated
locations on the image plane.
For each pixel in the source plane, we record the \emph{minimum}
intensity of the contributing image-plane pixels.
In this ``minimum-filtered'' map, any image-plane pixel that is blank
or low in intensity effectively ``vetoes'' any other contributions to
the source-plane intensity in that pixel.  For this reason, the
minimum map in fact represents (in the absence of noise and
pixelization effects) the only source-plane flux that can be
contributing to the residual map, consistent with the multiply imaging
lens potential model under consideration. This simple estimator
compensates for our inability to model simultaneously and effectively
all the features in the image, and focuses directly on the component
of the image we are interested in: the gravitationally-lensed
component.

When locating the best lens model for a given system, the scalar
objective function we compute is simply the total flux in the
minimum-filtered source plane.  In many cases we expect this 
scalar to
be zero, as the minimum filtering removes all unlensed flux: this is
certainly true of the blank parts of the image that 
are truly blank (rather
than containing sky noise). For this reason we ``de-noise'' the
galaxy-subtracted images by setting all pixels not containing
significant flux to zero. This thresholding is done with standard
object detection software \citep[SExtractor,][]{B+A96}, which
associates pixels together into objects and flags the remainder as
background.  Specifically, we use the ``objects'' checkimage output as
our data.

At each value of the model Einstein radii $\thetaE$, we maximize the
source plane flux by stepping through the external shear parameters.
This allows us to plot  source plane flux versus $\thetaE$. The
optimal model is then the one corresponding to the peak of this plot.

%%%%%%%%%%%%%%%%%%%%%%%%%%%%%%%%%%%%%
\begin{figure*}[t]
\centering\epsfig{file=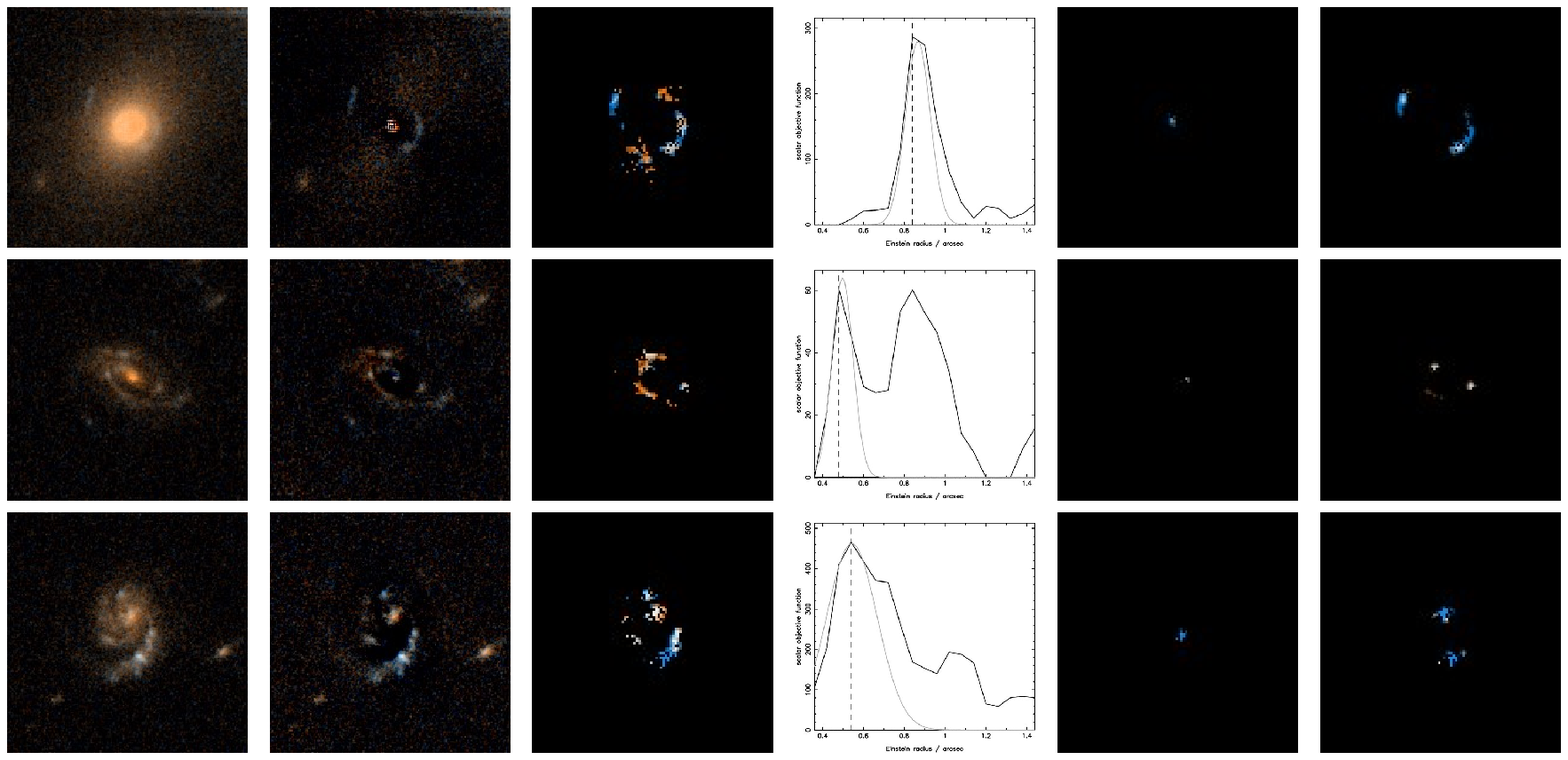,width=1.0\linewidth}
\caption{%
Robot modeling inputs and outputs for three example objects from the
EGS survey, one in each row of panels: Top row, the ``Anchor''
gravitational lens found by eye by \citet{Mou++07}; middle and bottom
rows, potential false positives, showing confusing lens plane spiral
arm structures, some of whose flux is consistent with having been
lensed.
In each row, numbering the panels from the left, we have: 1) the raw
\hst color cutout image; 2) the image after subtraction of a
Moffat-profile model for the lens galaxy light; 3) the thresholded
image input to the lens modeling robot; 4) the scalar objective
function (minimum-filtered source plane flux, see text) plotted as
function of Einstein radius and optimized over the external shear
parameters; 5) the optimal model source plane; 6) the corresponding
image plane. The images in panels 3 and 6 are used in computing the
goodness of fit statistic $\nsigma$ 
(Section~\ref{sect:method:classification:quality}).
Likewise, in panel 4 a Gaussian has been fitted to the peak of the 
scalar for
the purposes of estimating the uncertainty on the Einstein radius,
$\dthetaE$, and the detection ratio $\Rd$. All cutout images are
6~arcsec on a side.
\label{fig:robotexamples}}

\end{figure*}
%%%%%%%%%%%%%%%%%%%%%%%%%%%%%%%%%%%%%

% - - - - - - - - - - - - - - - - - - - - - - - - - - - - - - - - - - - - - -

\subsection{Multi-filter data}\label{sect:method:modeling:color}

So far we have made no mention of the achromatic nature of
gravitational lensing, yet this is one of the most important pieces of
information at the disposal of any by-eye lens searcher.  This is
because the human eye is very good at detecting subtle changes in
color, and in picking out objects of a given color from a noisy
background. However, the achromaticity is a direct result of surface
brightness being conserved at all wavelengths; up until this point
we have just described using surface brightness conservation in one
band. If imaging data in more than one band is available to us, how
should we include this information? We expect the source morphology to
be different between the different bands, so the lens inversions
should be done independently. However, it must be the same mass
distribution giving rise to any lensing effects detected.

The correct thing to do when analyzing the independent datasets (in
this case, the images in different filters) is to add the log
likelihoods for each band's image together. Dropouts (sources with no
detected flux in one of the bands) will rightly contribute zero to the
summed log likelihood. For computational efficiency we are not
optimizing the lens model by maximizing its likelihood, but we can
compute the likelihood of the ``optimal'' model once it has been
obtained. In Section~\ref{sect:method:classification} below we will do
exactly this -- and combine the likelihoods from the different filters
images as well. While multiple bands can be analyzed in this
straightforward way, we note that our method does not rely on having
multi-filter data -- although we expect performance to improve in the
multi-filter case.

One use of any additional filters could be 
in improving the pre-selection of candidate
elliptical galaxies.  Another is that the redder images tend 
to give
more accurate lens galaxy centroids (since the redder images have more
regular morphologies better tracing the dark matter halo).

% - - - - - - - - - - - - - - - - - - - - - - - - - - - - - - - - - - - - - -

\subsection{Modeling examples}\label{sect:method:modeling:examples}

In Figure~\ref{fig:robotexamples} we show three example bright red
galaxies (drawn from the EGS survey, see Section~\ref{sect:egs}), and
their automated analysis. One is a lens (visually selected by M07),
and two are potential false positives showing confusing ``lens plane''
structure. In each case the robot is able to find a lens model that
explains \emph{some} of the observed galaxy-subtracted residuals, but
with varying goodness of fit; in the next section we describe how we
use this and other information from the robot to make an automated
classification.

%---------------------------------------------------------------------------- 

\section{Automated lens candidate classification}\label{sect:method:classification}  

Given an optimized lens model, we now generate a set of data
describing the quality of the model in describing the observed image. 
A classification parameter can then be inferred from this 
robot output data vector~$\data$, whose components we describe below. 

% - - - - - - - - - - - - - - - - - - - - - - - - - - - - - - - - - - - - - -

\subsection{The robot's output: quantifying lens model quality}\label{sect:method:classification:quality}

The source plane flux scalar described above is simple, and fast, to
calculate. However, a much better-understood objective function is
the log likelihood of the predicted source model. To calculate this,
we invert the lens mapping to produce the corresponding image counts
predicted by the lens and source model,~$I^{\rm p}$. Assuming the
Gaussian approximation to the Poisson distribution for the (assumed
uncorrelated) pixel noise on the data image counts~$I$, the log
likelihood is given by
\begin{equation}
-2 \log L = \sum_i \frac{\left(I_i - b I_i^{\rm p}\right)^2}
                                    {\sigma_i^2}.
\label{eq:lhood} 
\end{equation}
The uncertainty $\sigma_i$ is estimated by adding in quadrature
the root mean square
pixel value measured in the background regions, to the square root of
the $i^{\rm th}$ detected pixel value (thus taking into
account, at least approximately, the Poisson noise due to the objects
themselves). We add a further term, also in quadrature, to attempt to
model the errors incurred by our very simple lens and coarsely-gridded
source models: since the inadequacies of our model are independent of
the quality of the data, we assume the modeling error will impose a
signal-to-noise ratio floor~$a$, so that its contribution to the
overall uncertainty can be (crudely) approximated by $I_i/a$. We find
that $a=4$ gives reasonable values of $\log L$ for a range of data
quality. In summary,
\begin{equation}
\sigma_i^2 = {\rm rms}^2 + I_i +  (I_i/a)^2.
\label{eq:errors} 
\end{equation}

The sum in equation~\ref{eq:lhood} is over all non-zero pixels in the
predicted image, since we are only attempting to explain the flux in
the multiply-imaged region around the putative lens. In principle, one
could use the presence of additional residual image structure that was
not compatible with being lensed as evidence that perhaps there is no
lensing occurring at all. However, we prefer at this stage to be
inclusive in the generated candidate list and only consider pixels in
the multiple-imaging regime.  The parameter $b$ is a rescaling factor
that allows us to correct the bias towards low source flux introduced
by the use of the minimum-filtered source plane intensity when
computing the predicted image~$I^{\rm p}$. We approximately
marginalize this parameter out by using the value of~$b$ that
maximizes $\log L$.

The left-hand side
of equation~\ref{eq:lhood} is distributed as chi-squared
for approximately $N$ degrees of freedom, where $N$ is the number of
non-zero pixels in the predicted image minus three model parameters.
Therefore, we record the goodness of fit as the number of standard
deviations~$\nsigma$ this scalar is from the mean of the chi-squared
distribution, using the Fisher approximation for large~$N$:
$\pr(2\chi^2|N) \approx G(2N-1,1)$ (where $G(m,w)$ is a Gaussian
distribution of mean~$m$ and standard deviation~$w$).  
The image-plane likelihood evaluation is time
consuming; however, as we use the source plane flux scalar to quickly
determine the ``best'' model, we need only compute~$\nsigma$ once.
The $\nsigma$ is the first component of the robot output data vector
$\data$. 

Our coarse parameter grid and simple lens model become increasingly
ill-suited to the data as the source size decreases.  For us to detect
any flux in the source plane the mass model must be good enough to map
the pixels back to the same point with sub-pixel precision. We are
helped by the PSF here, which spreads the flux around in the image
plane, but the resulting minimum source plane produces a very sparse
image plane.  We account for this in the likelihood evaluation by
applying a ``restoring beam'' to the predicted image. This beam is a
Gaussian with FWHM slightly less than that of the ACS PSF, and produces
more realistic image configurations. Since we originally mapped the
\emph{convolved} flux back to the source plane, we do the restoring
convolution such that the peak surface brightness of the source plane
is preserved into the predicted, convolved, image plane.

The parameter $\nsigma$ is clearly a powerful tool for quantifying how
good a fit to the data a lens model provides. However, it does not
tell the whole story. For example, very faint sources give very faint
fluctuations in the image plane, which can often easily be fitted
within the noise by a lens model. However, a human classifier is
unlikely to assign this candidate a high score.  We therefore make the
magnitude of the source the second component of $\data$. 
We also add two more
quantities. First is the uncertainty on the lens model Einstein
radius, $\dthetaE$: a convincing lens model should give a very small
$\dthetaE$. We estimate this from a Gaussian approximation
to the peak in the plot
of source plane flux against Einstein radius
(Figure~\ref{fig:robotexamples}, panel 4). The second is
related to this but subtly different: in a clean lens with just one
source plane, we expect only one lens model to match the data, and so
this same plot should have only one sharp peak. Therefore, as our
final $\data$-component we take the ratio~$\Rd$ of the integral under
the Gaussian fit to the integral under the whole source flux curve.

The four scalars described here that comprise $\data$ were chosen 
purely for
their intuitive information content. We are not suggesting they form a
complete set, or that they have been optimized for this application --
only that they are likely to provide some discriminatory power between
different candidates.  Indeed, we anticipate that much of the future
work improving the performance of the robot might center on generating
better $\data$-vectors, and so improving the ``capacity'' of the robot
to \emph{learn}. We will return to this issue in
Section~\ref{sect:discuss}.

%%%%%%%%%%%%%%%%%%%%%%%%%%%%%%%%%%%%%
\begin{table*}
\caption{Robot outputs for the three example lens candidates in
Figure~\ref{fig:robotexamples}.}
\label{tab:robotexamples}
\scriptsize
\begin{tabular}{cccccccccc}
\hline
Object name  & Position in                    & 
  $\nsigma$  & $\thetaE$  &  $\dthetaE$ & 
  $\Rd$  &  Source magnitude  & Human class & Robot class \\  
             & Figure~\ref{fig:robotexamples} & 
             & (arcsec)   &  (arcsec)   & 
         &  AB, F606W  & $(H)$ & $(H_r)$ \\ 
\hline\hline
HSTJ141833.11+524352.5 & top    & 4.11  & 0.84 & 0.06 & 0.62 & 28.40 & 3 & 2.9\\
HSTJ141856.16+523843.5 & middle & 8.93  & 0.48 & 0.05 & 0.27 & 30.23 & 0 & 2.1\\
HSTJ141828.06+523646.1 & bottom & 21.56 & 0.54 & 0.12 & 0.56 & 27.96 & 0 & 1.4\\
\hline
\end{tabular}
\end{table*}

%%%%%%%%%%%%%%%%%%%%%%%%%%%%%%%%%%%%%

We list the robot outputs $\data$ for the three examples shown in
Figure~\ref{fig:robotexamples} in Table~\ref{tab:robotexamples} --
this table illustrates some of the behavior suggested in
the above discussion.

% - - - - - - - - - - - - - - - - - - - - - - - - - - - - - - - - - - - - - -

\subsection{Inferring the human classification parameter}\label{sect:method:classification:inference}

Having defined our set of scalars describing the quality of a lens
model (the robot output~$\data$), we must now decide on our criteria
for assessing the robustness of each strong lens candidate.  Before we
continue, let us try to understand this last statement. Naively we
might hope to quantify the quality of a lens candidate by seeking the
``probability that it is a gravitational lens.''  However, to
calculate this it would also be necessary to compute the probability
that the candidate is NOT a gravitational lens as well. The problem
is that we do not yet have a sufficiently detailed and quantitative
understanding of galaxy morphology to be able to do this. In the
presence of disk components, spiral arms, satellite galaxies and so
on, it is genuinely difficult to disentangle the contributions to the
image from the lensed sources versus from the contribution from the candidate
galaxy itself. Clearly, if the value of the detected source plane flux
is zero then the system can be rejected. But we can expect a further
continuum of likelihood values: if the robot is to succeed in
returning a useful, shortened candidate list for our inspection, it
must first learn what makes a good candidate.

While our quantification of galaxy structure is poor, our ability to
identify gravitational lenses in high resolution images by eye is
not. This suggests that a much better-defined measure of 
lens candidate
robustness is the posterior probability distribution
$\pr(H|\data)$, where $H$~is the \emph{classification that a human
  would have assigned the system}.  

In Table~\ref{tab:classsystem} we propose a simple four-point system
for human classification of gravitational lenses. Our experience is
that fewer than four classes is not flexible enough to describe a set
of lens candidates, while the differences between any more than four
classes become too small for different classifiers to agree
upon. 

%%%%%%%%%%%%%%%%%%%%%%%%%%%%%%%%%%%%%
\begin{table}
\caption{Human lens classification system used in this work.}
\label{tab:classsystem}
\begin{tabular}{cl}
\hline
Class $H$  & Meaning \\  
\hline\hline
0 & Definitely not a lens \\
1 & Possibly a lens  \\
2 & Probably a lens  \\
3 & Definitely a lens  \\
\hline
\end{tabular}
\end{table}

%%%%%%%%%%%%%%%%%%%%%%%%%%%%%%%%%%%%%

If we have a set of lens candidates, each with robot data vector
$\data_i$ and \emph{known human class $H_i$}, then we can estimate
the probability density function (PDF)
$\pr(\data|H)$ from these clouds of points in $\data$-space, $H$-value
by $H$-value. We can then use this model PDF to compute
$\pr(\data_j|H)$ for the $j^{\rm th}$ candidate. $\pr(H|\data_j)$ is
given by Bayes' Theorem, 
\begin{equation} 
  \pr(H|\data_j) \propto  \pr(\data_j|H)\pr(H). 
\end{equation}
The prior PDF $\pr(H)$ encodes our expectations for the relative
frequencies of each human classification $H$. One can imagine that
there might be a great many low-$H$ (poor quality) candidates and only
a few high-$H$ (and so very robust) candidates. In
Section~\ref{sect:training:class} below 
we will explore two quantified prior PDFs.

To make progress we now need to determine $\pr(\data|H)$.  
For this we
require a large sample of candidates, containing both lenses and
non-lenses, and showing similar problems to real candidates. In the
following subsection we describe the two sources of this 
\emph{training set}, and the resulting education of our robot.

%-------------------------------------------------------------------------------

\section{Training the robot}\label{sect:training}

Having implemented an efficient lens modeling robot, we now have to
transfer to it our knowledge of lens, and more importantly, intrinsic
BRG structure.  The procedure is to generate a well-defined set of
lens candidates whose human classification~$H$ is known, and that
sample well the possible range of $H$-values.

% - - - - - - - - - - - - - - - - - - - - - - - - - - - - - - - - - - - - - -

\subsection{Non-lenses in the EGS survey}\label{sect:training:egs}

The 63 ACS images\footnote{Available from {\tt
    http://aegis.ucolick.org}} of the
0.18\,deg$^2$ \hst mosaic of the Extended Groth Strip survey
\citep[EGS,][]{Dav++07} were taken in both the F814W (``$I_{\rm
  814}$-band'') and F606W (``$V_{\rm 606}$-band'') filters, to depths
of 28.14 and 27.52\,mag respectively (5-$\sigma$ limits for point
sources in 0.12\,arcsec-radius circular measurement apertures). This
survey contains a control sample of three strong lenses confirmed by
their image modeling (and in two cases by spectroscopy as well), 
and four
plausible strong lens candidates, 
all identified by an independent search by
visual inspection of the ACS frames~(M06 \& M07). We will use this
sample in Section~\ref{sect:egs} for comparison with our automated
search results. Here, we aim to produce a catalog of BRGs that
includes these known lenses, and then define a sample of known
\emph{non-lenses} with which to educate the robot.

To avoid any morphological bias in the candidate lens galaxies, we
only minimally pre-filter the catalog of objects detected in the EGS
fields, applying only a signal-to-noise and a color cut. Objects with
$I_{\rm 814} \le 22.0$ and $(V_{\rm 606} - I_{\rm 814}) > 0.8$ were
selected, giving a sample of 1085 BRGs.  All of the confirmed lenses, and 
three of the four lower-quality 
lens candidates, identified by M06 \& M07 passed this cut. For our
training set of non-lenses, we divided the 63 EGS fields into two
sets, a training set of 3 fields, and a survey set of 60 fields. The
three training fields were chosen to contain none of the lens
M06/M07 candidates: the fields chosen were
\texttt{egs2101}, \texttt{egs2102}, and \texttt{egs2103}. We then
analyzed the 97 BRGs in these three training fields that passed our
color-magnitude cut, generating optimized lens models for each
object. These systems were classified by one of us (PJM), and binned
by $H$-value.

% - - - - - - - - - - - - - - - - - - - - - - - - - - - - - - - - - - - - - -

\subsection{Simulated lenses}\label{sect:training:sim}

Sampling BRGs that are lenses is harder -- strong gravitational lenses
are rare objects. Many of the known galaxy-scale lenses were found
either via source-oriented radio source or quasar
surveys~\citep[\eg][]{Bro++03,Ogu++06}, or in unresolved spectroscopic
surveys~\citep[\eg][]{Bol++06}; they tend to have either point-like or
highly-magnified images, and are therefore not representative of the
lenses we expect to find in high resolution imaging surveys.

% The largest sample of strong
% galaxy-galaxy lenses is that of the SLACS survey~\citep[][Bolton et al.\ in
% preparation]{Bol++06}. However, since these lenses were selected from the SDSS
% spectroscopic survey, the median lens redshift is $\sim0.2$  and the
% magnification bias is rather high, leading to a rather high fraction of high
% magnification ring-like systems around very bright BRGs.

The best training set for our purposes is then a sample of lenses
whose properties match those of the lenses we expect to find in a high
resolution optical imaging survey. To this end we simulated a
realistic population of lenses and generated mock images of them. For
the lenses, we used a set of 68 morphologically-selected early-type
galaxies from the EGS survey ACS images. We made a very rough estimate
of their redshift, by assuming a velocity dispersion of 220\,km/s and
using the Faber-Jackson relation to estimate luminosity, and so
distance modulus given apparent $I_{814}$-band magnitude. We then drew
a velocity dispersion from a Gaussian distribution of mean 220\,km/s
and width 20 km/s. The variety of redshifts and velocity dispersions
is sufficient to give a reasonable distribution of Einstein radii. We
also measured the position, ellipticity and orientation of the
early-type galaxy light: these parameters were then used, along with
the model velocity dispersion, to define an SIE model lens for each
lens galaxy. To complete the lens model we added external shear with
amplitude drawn from a log-normal distribution of mean 0.05 and width
0.6 \citep{H+S03}, and external convergence equal to half the external
shear magnitude.

For the sources, we drew faint galaxies from the EGS catalog,
extracting their magnitudes, sizes, ellipticities and orientations and
used these data to define a simple \citet{deV48} bulge plus
exponential disk model. We made a robust faint galaxy selection
(\mbox{$22 < I_{\rm 814} < 27$}, and
\mbox{$0.15\thinspace\mathrm{arcsec} < {\rm FWHM} <
0.36\thinspace\mathrm{arcsec}$}) and approximately corrected the sizes
for the ACS PSF by subtracting its Gaussian-approximated 
width in quadrature; the resulting corrected 
half-light radius was then asserted for both components of 
the disk+bulge
model. We drew the disk/bulge ratio from a Gaussian of mean 3.0 and
width 0.5, since we expect most faint blue source galaxies to be
disk-like. For the source redshift we again approximated it using the
$I_{814}$ magnitude, following a simplified version of the recipe used
by \citet{Mas++04b}: $\pr(z_s)$ was assumed to be a Gaussian of width
0.4, centered on $z_m=0.7+0.15\,(I_{814}-22)$. 

We generated a number of possible sources for each lens galaxy, 
providing more lens systems than early-type galaxies. Once we had
both source and lens redshift, and SIE velocity dispersion parameter,
we computed the Einstein radii and rejected systems with $\thetaE <
0.4$\,arcsec as being unobservable
(Section~\ref{sect:method:modeling:technical}). We then generated a
source position drawn uniformly from a box of width $0.3\thetaE$
centered on a point offset by $(0.15,0.15)\thetaE$ from the optical
axis of the lens. This rather complex recipe was designed to avoid an
overabundance of ring-like lenses while keeping us within the
reasonably high
magnification regime. Finally, we shuffled the list of simulated lens
systems and selected the first 100 systems to make simulated images.

%%%%%%%%%%%%%%%%%%%%%%%%%%%%%%%%%%%%%
\begin{figure*}[t]

\begin{minipage}[t]{\linewidth}
\begin{minipage}[t]{0.48\linewidth}
\centering\epsfig{file=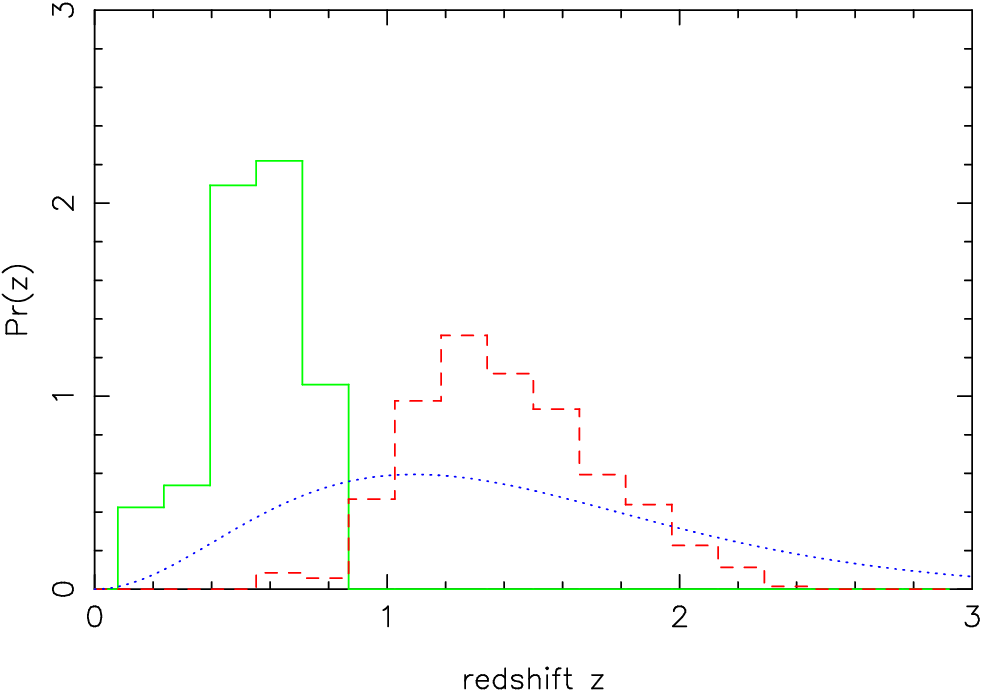,width=1.0\linewidth}
\end{minipage} \hfill
\begin{minipage}[t]{0.48\linewidth}
\centering\epsfig{file=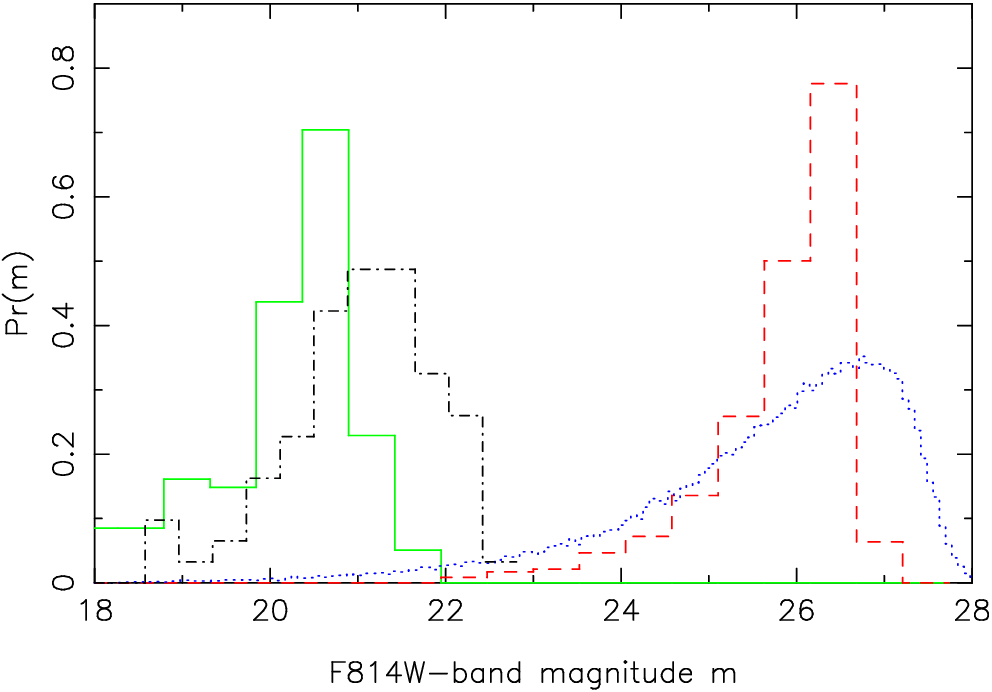,width=1.0\linewidth}
\end{minipage}
\end{minipage}

\caption{%
Redshift (left) and $I_{814}$-band magnitude (right) distributions for
lens and source galaxies in the 100 simulated lens systems used in
this paper.  Green solid histograms show simulated lens galaxies, red
dashed histograms show simulated source galaxies. Blue dotted curves
show the parent source population -- the redshift distribution from
\citet{Lea++07} and the EGS source counts. The black dot-dashed
histogram is the morphologically-selected spheroid sample from
\citet{Tre++05}.}

\label{fig:simdists}

\end{figure*}
%%%%%%%%%%%%%%%%%%%%%%%%%%%%%%%%%%%%%

Figure~\ref{fig:simdists} shows the redshift and magnitude distributions of
the simulated lenses. The source redshifts can be seen to be roughly
consistent with having been drawn from a distribution like that derived by
\citet{Lea++07} for the appropriate EGS magnitude limit, 
once the lensing (which favors higher source redshifts) has
been taken into account. The magnitudes of the lenses are consistent with the
bright (and hence massive) end of the early type population \citep[as
characterized using the morphologically-selected and
spectroscopically-observed GOODS sample of][]{Tre++05}. The (unlensed)
magnitudes of the sources are consistent with their parent EGS population,
with the piling-up at magnitude 26 being due to a combination of the magnitude
limit and the lensing shift to higher mean redshift.

%%%%%%%%%%%%%%%%%%%%%%%%%%%%%%%%%%%%%
\begin{figure}[t]
\centering\epsfig{file=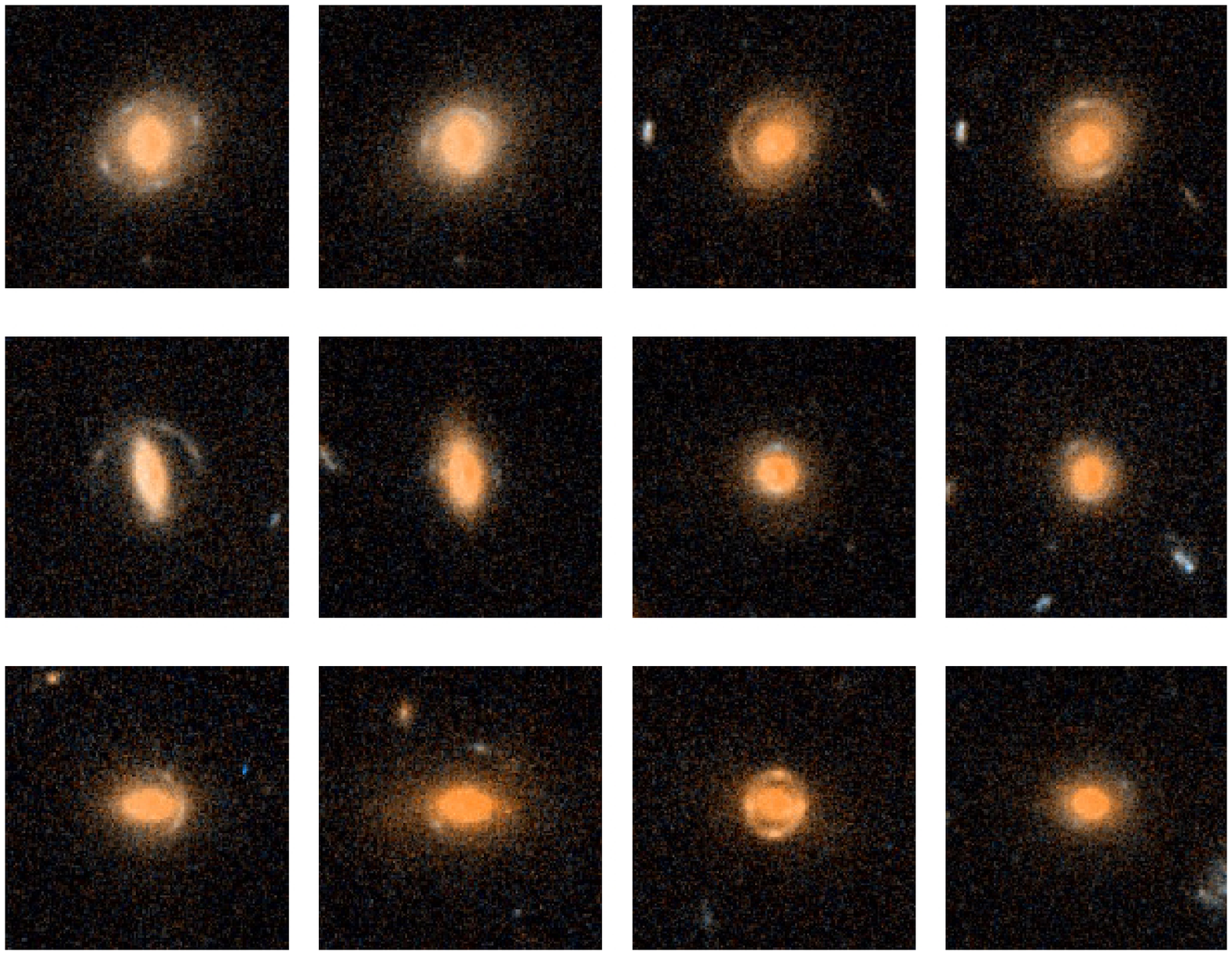,angle=90,width=1.0\linewidth}
\caption{%
Mock EGS data for 12 example simulated gravitational lenses.
All cutout images are 6\,arcsec on a side.} 
\label{fig:sims}
\end{figure}
%%%%%%%%%%%%%%%%%%%%%%%%%%%%%%%%%%%%%

We made mock, noise-free ACS images of the lensed images given the
lens model described above; the process is the same as described
(albeit in an inferential setting) in M07 and \citet{Mar++07}. We
approximated the ACS PSFs by a Gaussian of FWHM\,=\,0.14\,arcsec and
convolved the simulated images with this kernel, before adding it to
the relevant EGS early-type galaxy image (in each filter). The
resulting composite image has approximately the correct noise
properties: we neglected the small contribution to the noise from the
lensed images, as the lens galaxy surface brightness is often higher
anyway.  The end result is a set of 100 simulated lens galaxy cutout
images that can be analyzed in an identical manner to the EGS lens
candidates themselves. A gallery of examples is given in
Figure~\ref{fig:sims}.

% - - - - - - - - - - - - - - - - - - - - - - - - - - - - - - - - - - - - - -

\subsection{Modeling $\pr(\data|H)$}\label{sect:training:pdf}

We divided our training set (100 simulated lenses, plus 97 non-lenses
from EGS) into 4 $H$-value bins, and for each $H$ bin, plotted all of
the
robot output parameters in the vector~$\data$ 
against each other. These parameters
were assumed to have been drawn from the distribution $\pr(\data|H)$;
from the plots we estimated the number of modes of this PDF, and
selected the samples associated with each mode by a series of
orthogonal cuts in the values of $\data$. The modes were then modeled
as multivariate Gaussians, defined by the means and covariance matrix
of each mode. After normalizing each mode by the number of samples
used to compute the properties of the mode, the overall PDF
$\pr(\data|H)$ for that value of $H$ is given by the simple sum of
these Gaussian modes.

In some cases the modes were broadened to provide a more satisfactory
picture of the PDF; in some cases the Gaussian approximation breaks
down leaving an ill-fitting distribution. This is of course a very
subjective way of modeling the PDF, but this does not matter: all we
are attempting to do is transfer our expertise to the robot such that
it agrees with us in the generation of $H$-values. Since these are
themselves subjective there seems little point in insisting on
objectivity at every turn. While we can see there is room for
improvement in the accuracy of the PDF model, the robot's more
important quality is that of reproducibility: having taught it about
human classification, the robot will be consistent in its own
classification (unlike humans), such that a selection function can be
estimated from realistic simulated data.

%%%%%%%%%%%%%%%%%%%%%%%%%%%%%%%%%%%%%
\begin{figure*}[!ht]
\begin{center}
\begin{minipage}{0.9\linewidth}
\begin{minipage}{0.48\linewidth}
\centering\epsfig{file=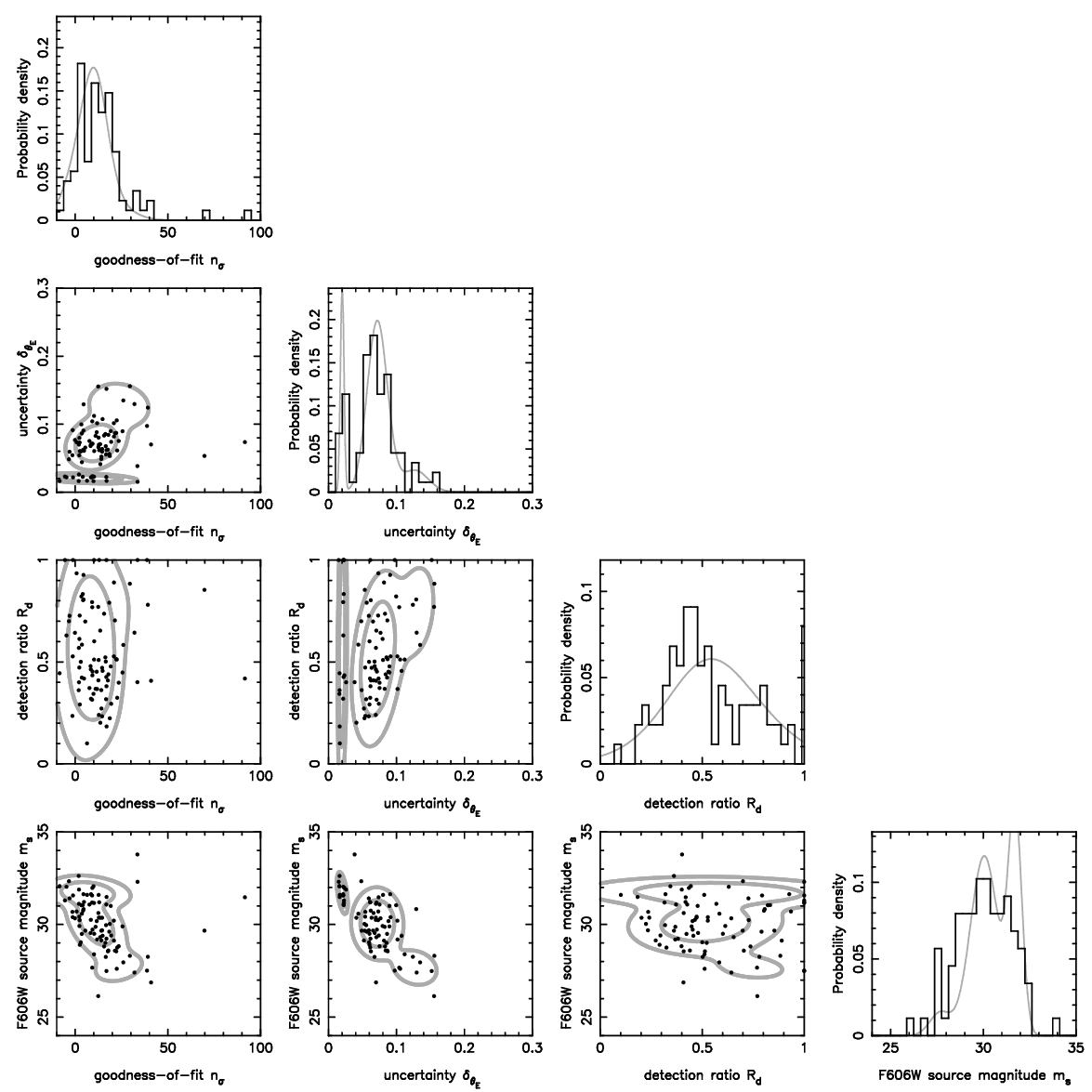,width=1.0\linewidth}
\caption{%
$\pr(\data|H=0)$ derived from the robot outputs (points) for the
  training set. Points correspond to objects classified by a human
  (PJM) as class $H=0$, definitely not a lens. In this and subsequent
  PDF figures the contours enclose
  68\% and 95\% of the total probability.   
\label{fig:pdf0}}
\end{minipage}\hfill
\begin{minipage}{0.48\linewidth}
\centering\epsfig{file=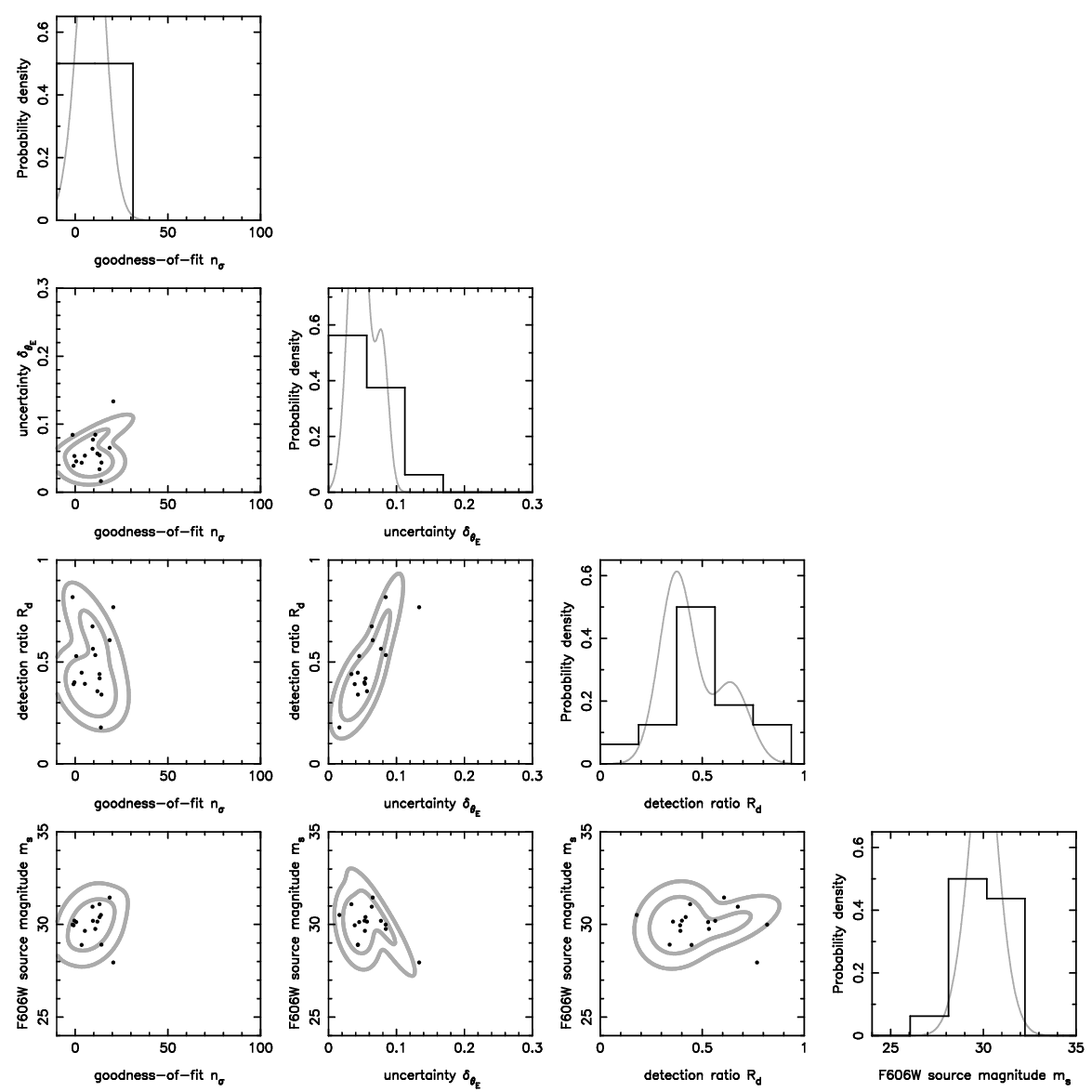,width=1.0\linewidth}
\caption{%
$\pr(\data|H=1)$ derived from the robot outputs (points) for the
  training set. Points correspond to objects classified by a human
  (PJM) as class $H=1$, possibly a lens.    
\label{fig:pdf1}}
\end{minipage}
\end{minipage}
\end{center}
\vspace{-2\baselineskip}
\begin{center}
\begin{minipage}{0.9\linewidth}
\begin{minipage}{0.48\linewidth}
\centering\epsfig{file=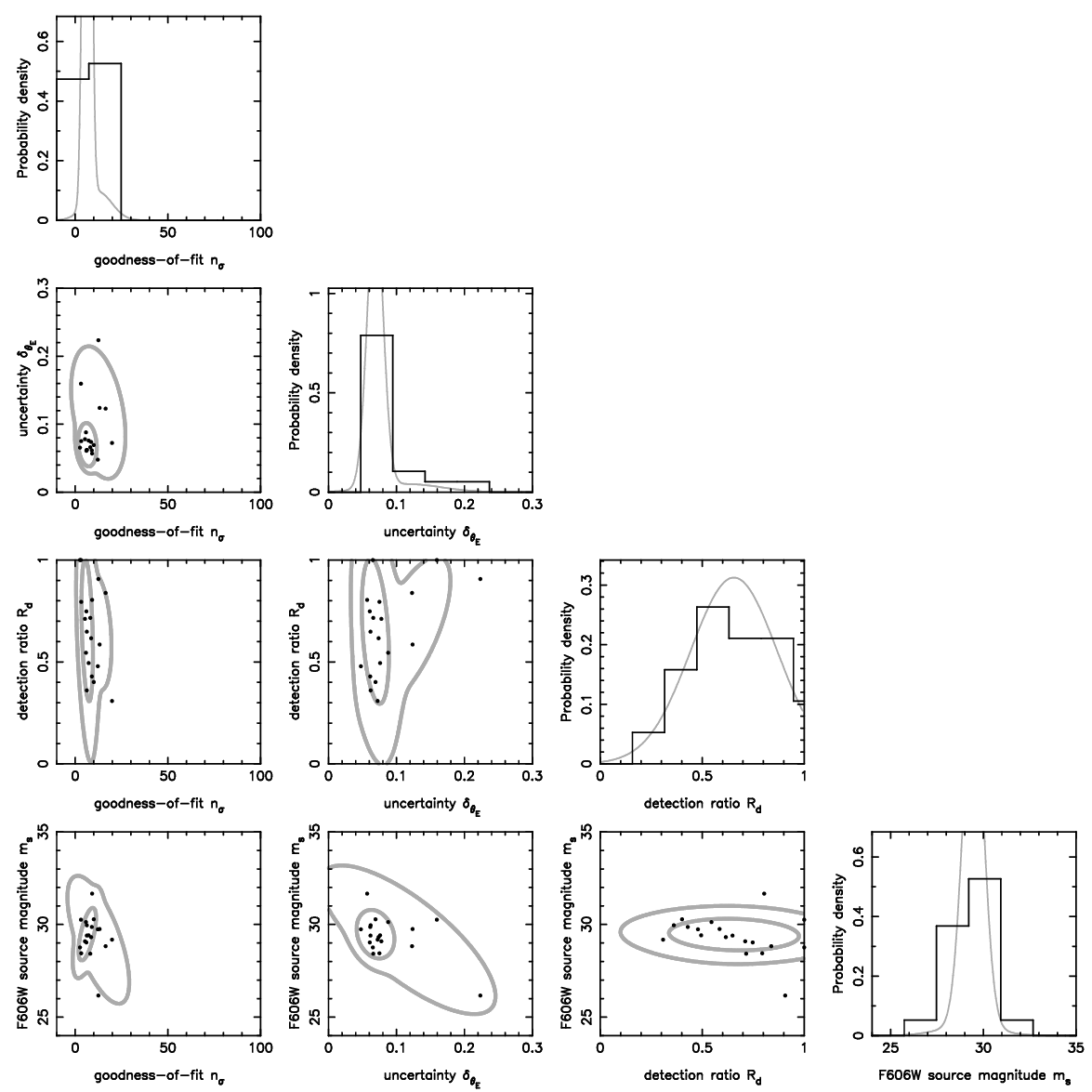,width=1.0\linewidth}
\caption{%
$\pr(\data|H=2)$ derived from the robot outputs (points) for the
  training set. Points correspond to objects classified by a human
  (PJM) as class $H=2$, probably a lens.   
\label{fig:pdf2}}
\end{minipage}\hfill
\begin{minipage}{0.48\linewidth}
\centering\epsfig{file=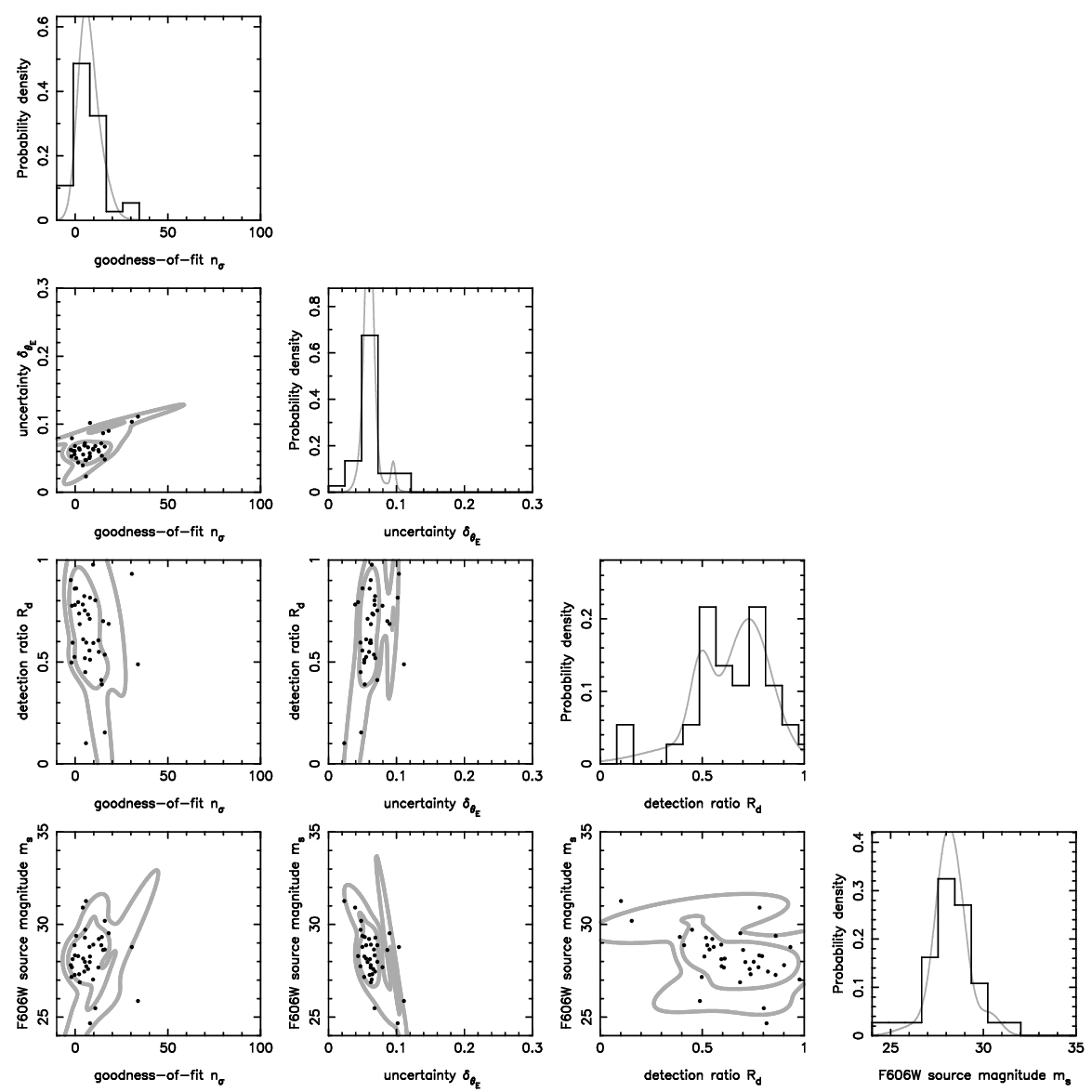,width=1.0\linewidth}
\caption{%
$\pr(\data|H=3)$ derived from the robot outputs (points) for the
  training set. Points correspond to objects classified by a human
  (PJM) as class $H=3$, definitely a lens. 
\label{fig:pdf3}}
\end{minipage}
\end{minipage}
\end{center}
\end{figure*}
%%%%%%%%%%%%%%%%%%%%%%%%%%%%%%%%%%%%%

In Figures~\ref{fig:pdf0}--\ref{fig:pdf3} we show plots of all four
PDFs modeled in this way: $\pr(\data|H=0)$, $\pr(\data|H=1)$,
$\pr(\data|H=2)$, and $\pr(\data|H=3)$.  These plots provide the
quantification of the common sense we would like the robot to have
when inspecting lens candidates. In each panel, the contours enclose
68\% and 95\% of the marginalized probability density.

% - - - - - - - - - - - - - - - - - - - - - - - - - - - - - - - - - - - - - -

\subsection{Robotic lens classification}\label{sect:training:class}
% if (keyword_set(NAIVE)) then begin
%   Prior = [1,1,1,1]/4.0
% endif else if (keyword_set(OPTIMISTIC)) then begin
%   Prior = [50,100,250,600]/1000.0
% endif else if (keyword_set(DECISIVE)) then begin
%   Prior = [450,50,50,450]/1000.0
% endif else begin
%   Prior = [900,80,19,1]/1000.0
% endelse

In this Section we describe the automated classification of the
training set, and estimate the completeness and purity 
of the resulting sample. 

We
use the PDFs described in the previous Section to compute the
posterior PDF for the classification parameter~$H$ as follows:
\begin{equation}
\pr(H|\data) \propto \sum_i \pr(\data|H=i)\pr(H=i),
\end{equation}
and then choose as our estimator for $H$ the posterior mean, denoted
$\Hr$ (where ``r'' is for ``robot''): 
\begin{equation}
\Hr = \frac{\sum_i i\; \pr(H=i|\data)}{\sum_i \pr(H=i|\data)}.
\end{equation}
We note that $\Hr$, the robot's estimate of $H$, is a continuous
variable even though $H$ is not. In Section~\ref{sect:egs} below we
combine human classification parameters from several human inspectors,
and so there $H$ does not take integer values either.

In order to infer the human classification~$H$ we need to specify the
prior probability~$\pr(H)$. This is exactly what the human classifier
is doing when reminding herself that lenses are rare, so that class-3
candidates should be considerably 
rarer than class-0 objects (which make up the
majority of objects). In this spirit we might assign a prior based on
what we expect the fractions of the different classes might be. A
reasonable estimate for the number of BRGs acting as lenses is 1 in
1000, and we might hope for, say, 20 times more probable lenses than
actual lenses. With a similar argument for class 1, we end up with
$\pr(H) = [0.9,0.08,0.019,0.001]$. 

\begin{figure*}[!ht]
\begin{minipage}[t]{0.48\linewidth}
\begin{minipage}[t]{0.48\linewidth}
\centering\epsfig{file=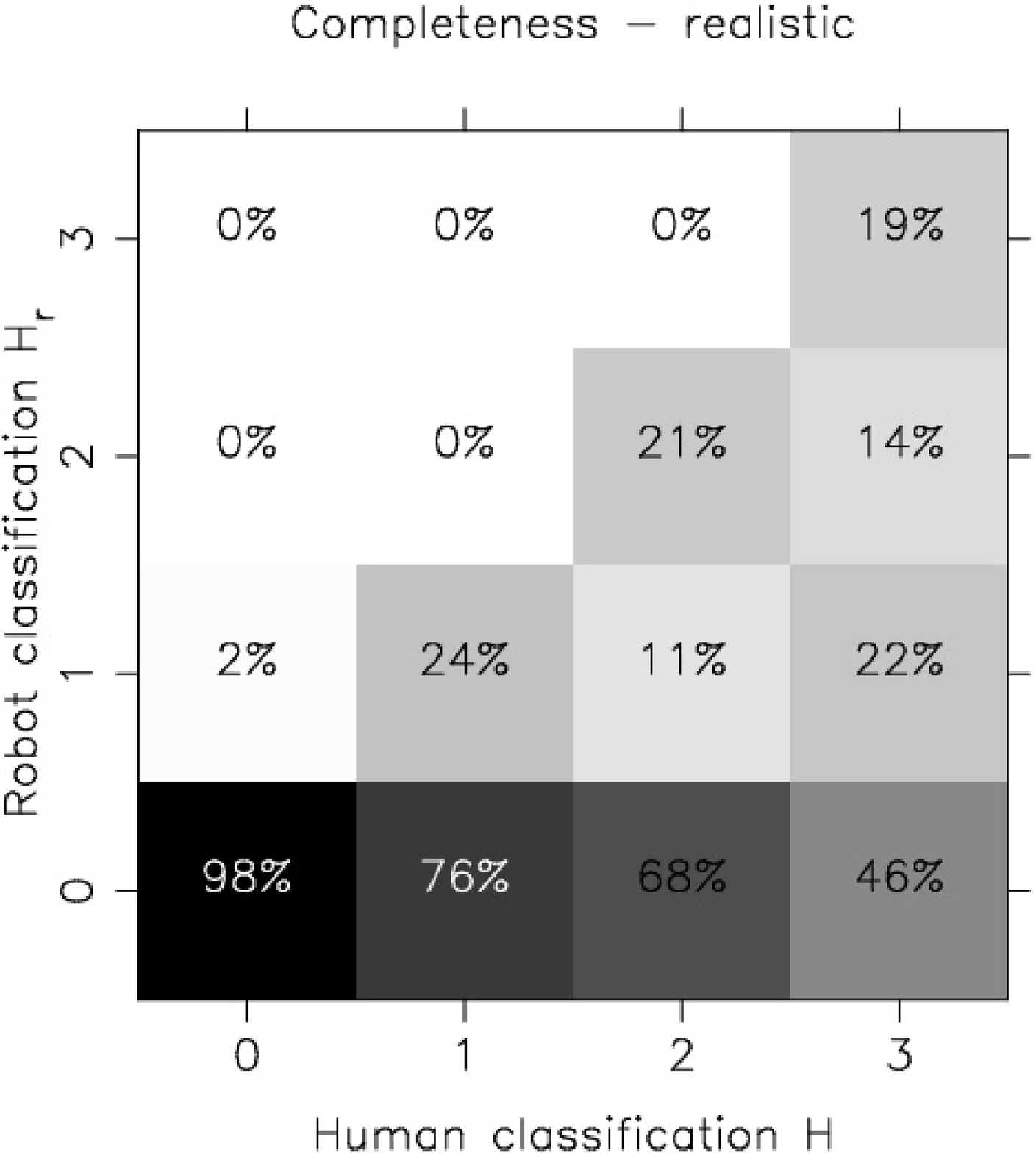,width=1.0\linewidth}
\end{minipage} \hfill
\begin{minipage}[t]{0.48\linewidth}
\centering\epsfig{file=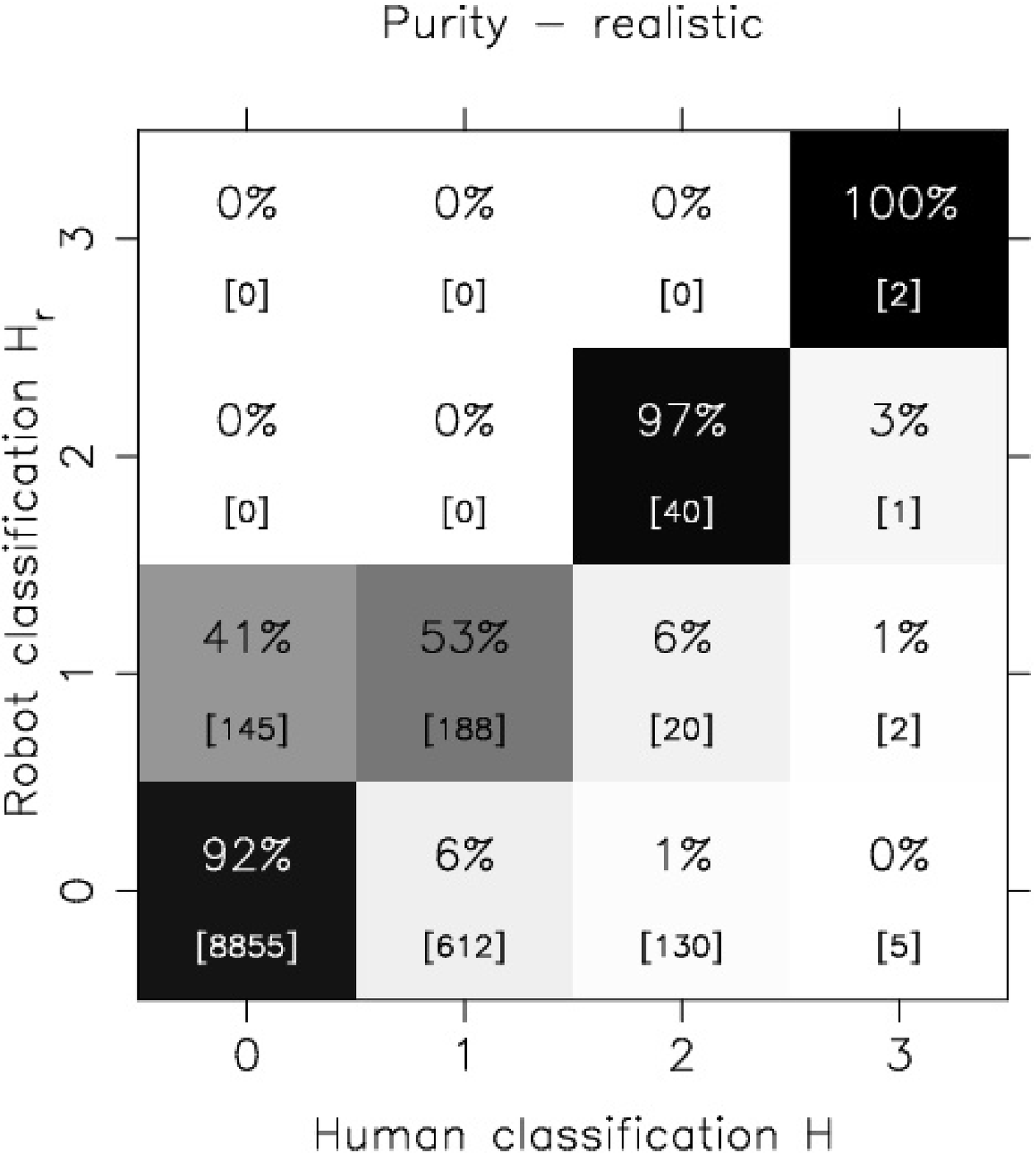,width=1.0\linewidth}
\end{minipage}
\end{minipage} \hfill
\begin{minipage}[t]{0.48\linewidth}
\begin{minipage}[t]{0.48\linewidth}
\centering\epsfig{file=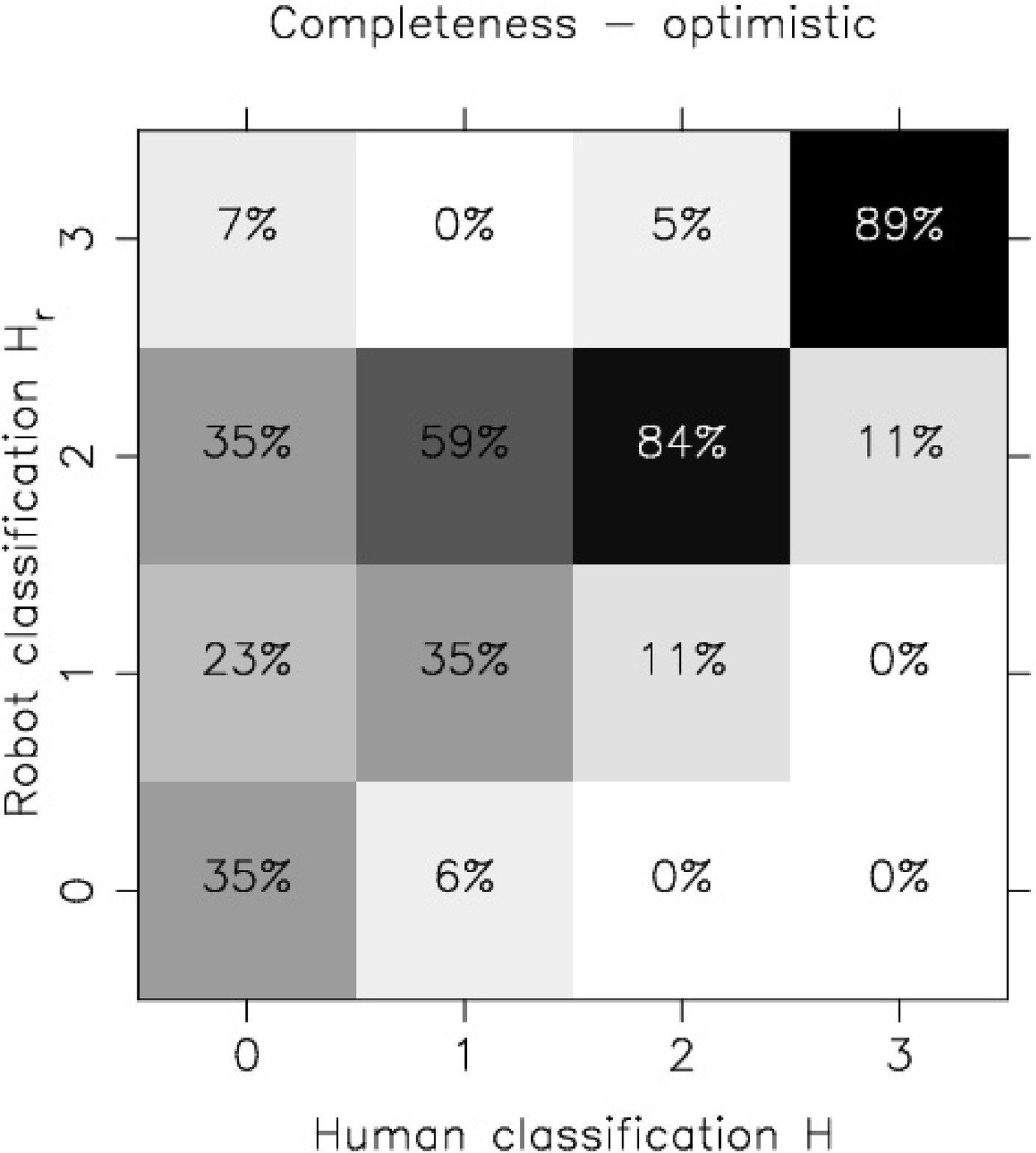,width=1.0\linewidth}
\end{minipage} \hfill
\begin{minipage}[t]{0.48\linewidth}
\centering\epsfig{file=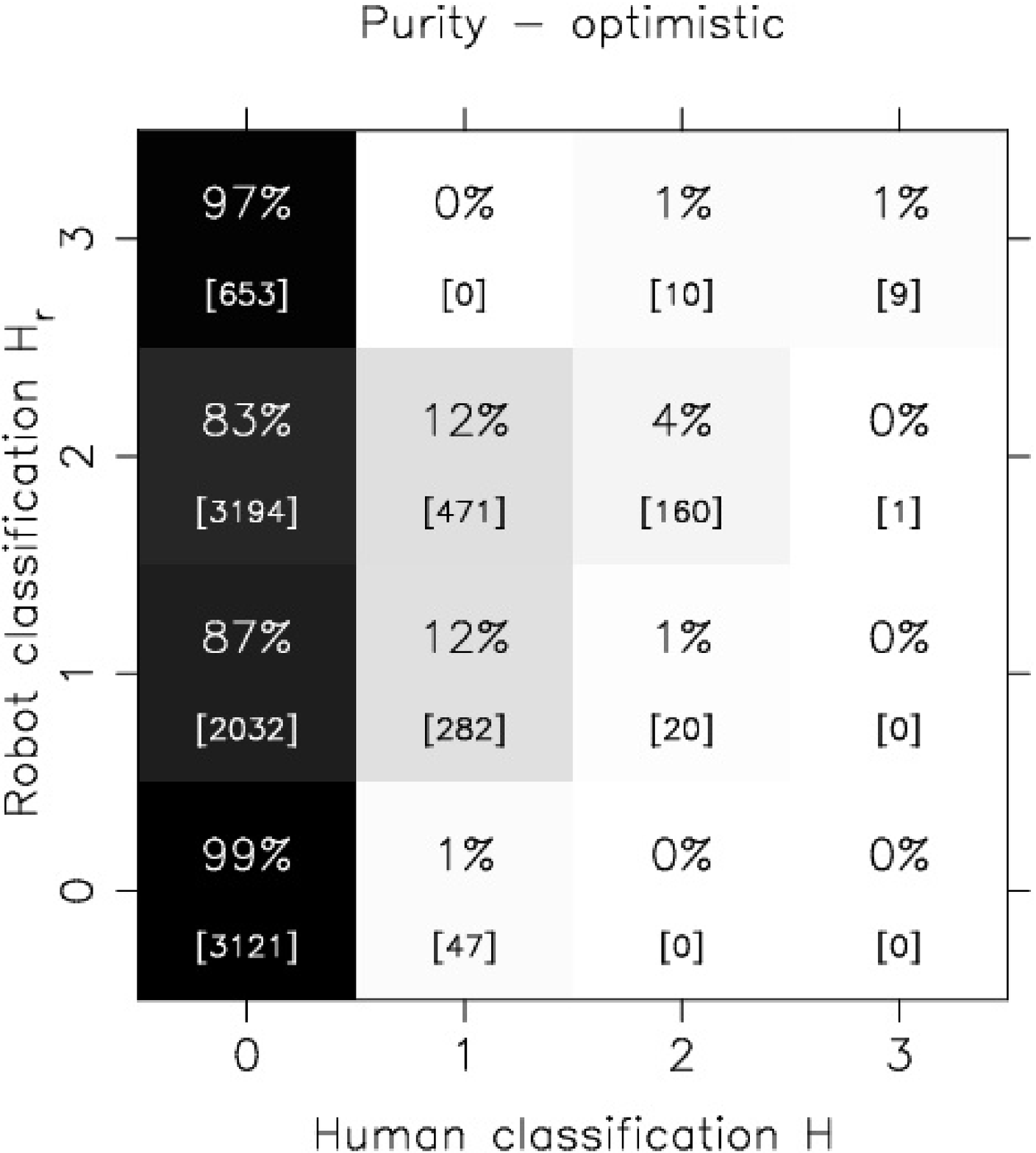,width=1.0\linewidth}
\end{minipage}
\end{minipage}
\caption{%
Left two panels: completeness (left) and purity (right) charts for a
{\it realistic} prior distribution of $H$-values, namely one where the
object classifications are as expected given what we know about the
relative scarcity of gravitational lenses.  Right two panels: same,
but for an {\it optimistic} prior distribution of $H$-values, namely
one where the majority of BRGs are expected to be acting as observable
gravitational lenses.
Completeness values are the percentages of the total number of
candidates of true class~$H$ in each robot class~$\Hr$ bin -- these
percentages sum to 100 in columns.  Purity values are the percentages
of the total number of candidates in each robot class~$\Hr$ bin that
have true class~$H$ -- these percentages sum to 100 in rows.  The
numbers in parentheses on the purity charts are the approximate
expected numbers of objects in each bin for a 1~square-degree survey.
\label{fig:completenesschart:istic}}
\end{figure*}

With this assumption, we calculated the posterior mean $\Hr$ for each
object in the training set, and compared it with its true
(human-determined) value of $H$. This comparison is shown as a
``completeness chart'' in the far left-hand panel of
Figure~\ref{fig:completenesschart:istic}. Completeness $C(H;\Hr)$ is
defined as the percentage of objects with human class~$H$ that have
robot class~$\Hr$. For example, in
Figure~\ref{fig:completenesschart:istic}, 19\% of objects with human
class~$H=3$ were found to also have robot-assigned class~$\Hr=3$. In
this chart, the percentages sum to 100 in columns. The ideal robot
would give a completeness chart that would be white (0\%) everywhere
except on the diagonal, where it would be black (100\% complete).

The apparently low success rate of the robot at high-$H$ must be
measured against its performance at low-$H$: an impressive 98\% of
definite non-lenses are rejected. This is just the usual trade-off
between completeness and purity. Defining purity $P(\Hr;H)$ as the
percentage of objects with robot class~$\Hr$ that actually have human
class~$H$, we can plot this as a complementary purity chart -- except
that now, the proportions of objects in each human class $H$-bin need
to be realistic for the numbers to make sense.  Our training set does
not have realistic proportions of objects of each class (nearly half
are known to be lenses); this is fine for estimating completeness,
but must be corrected when estimating purity. Here we simply adjust
the calculated purities to what they would have been had the
proportions been those assumed in the prior $\pr(H)$ defined above. 
In the purity
chart, the percentages sum to 100 in rows. The ideal robot would give
a purity chart that would be white (0\%) everywhere except on the
diagonal, where it would be black (100\% pure).  We show the purity
chart for the realistic-prior robot in the second left-hand panel
Figure~\ref{fig:completenesschart:istic}. 

%%%%%%%%%%%%%%%%%%%%%%%%%%%%%%%%%%%%
%%%%%%%%%%%%%%%%%%%%%%%%%%%%%%%%%%%%%

High purity means high efficiency: with a realistic prior for the
abundance of each class of candidates, the robot class 3 sample is
100\% pure. This is certainly very encouraging for future surveys,
where the human inspection is time consuming and costly. However, in
the short term we might prefer a relatively impure sample of
candidates to be output by the robot, in return for a higher
completeness. We can achieve this by altering the prior on~$H$,
analogous to having a inspector of different {\it character}.  An
``optimistic'' robot might have a prior PDF $\pr(H) =
[0.05,0.10,0.25,0.60]$, which, while not reflecting our expectations
of the abundance of lenses at all, has the practical effect of 
increasing the chances of a high
classification value. In the rightmost two panels of
Figure~\ref{fig:completenesschart:istic} we show the completeness and
purity of the samples generated by a robot of this character -- its
behavior is now such that the purity of the $\Hr=3$ bin is low (1\%),
while the completeness at $H=3$ is rather high (89\%). The optimistic
robot's classifications for the example systems in
Figure~\ref{fig:robotexamples} are shown in the final column of
Table~\ref{tab:robotexamples}.

\begin{figure}[!h]
\begin{center}
\begin{minipage}[t]{0.9\linewidth}
\begin{minipage}[t]{0.48\linewidth}
\centering\epsfig{file=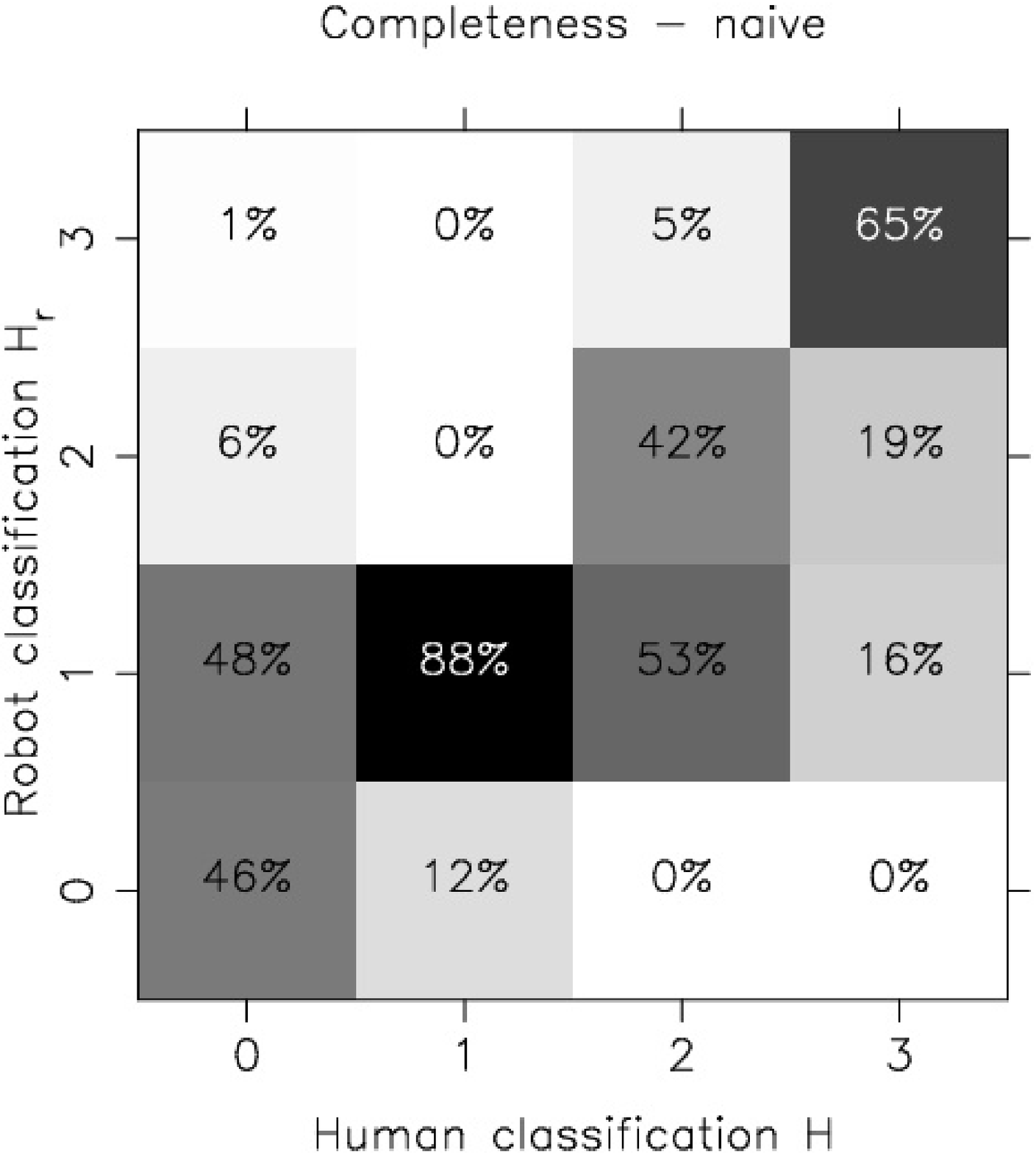,width=1.0\linewidth}
\end{minipage} \hfill
\begin{minipage}[t]{0.48\linewidth}
\centering\epsfig{file=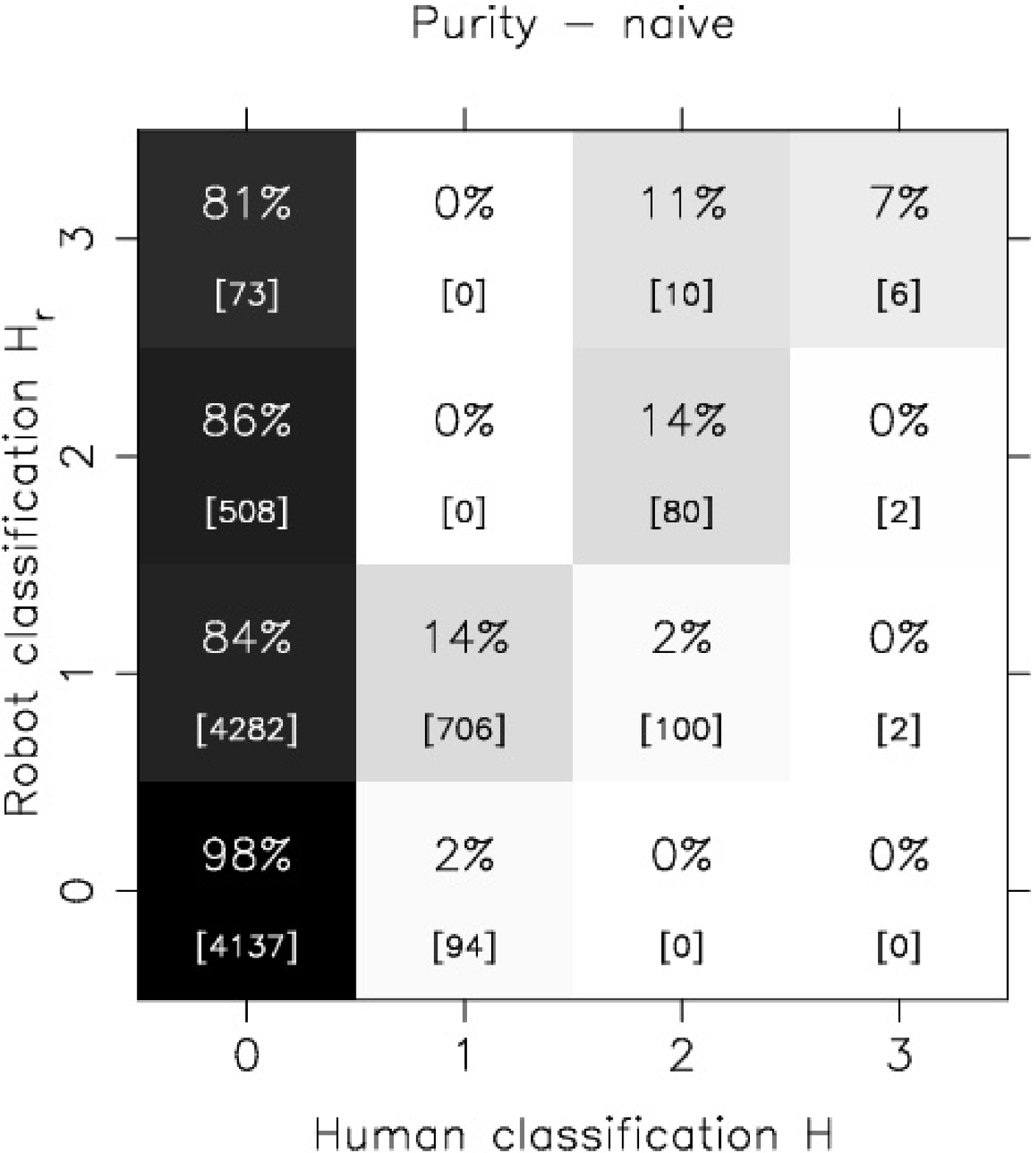,width=1.0\linewidth}
\end{minipage}
\end{minipage}
\end{center}
\caption{%
Completeness (left) and purity (right) charts for a {\it naive} prior
distribution of $H$-values, namely one where all classifications are
considered equally likely to occur.
\label{fig:completenesschart:naive}}
\end{figure}

Finally, for comparison we also show the performance of a ``naive''
robot -- one whose prior is uniform in $H$ -- in
Figure~\ref{fig:completenesschart:naive}. While this is something of a
compromise between optimism and realism, it is not designed to be
such; it is neither complete enough nor pure enough to be useful. This
should not be a surprise: the prior associated with this robot is one
of maximal ignorance. We summarize the
prior PDFs associated with each robot character in
Table~\ref{tab:character}.

%%%%%%%%%%%%%%%%%%%%%%%%%%%%%%%%%%%%%
\begin{table}[!h]
\caption{Robot characters and corresponding prior PDFs $\pr(H)$.}
\label{tab:character}
% if (keyword_set(NAIVE)) then begin
%   Prior = [1,1,1,1]/4.0
% endif else if (keyword_set(OPTIMISTIC)) then begin
%   Prior = [50,100,250,600]/1000.0
% endif else if (keyword_set(DECISIVE)) then begin
%   Prior = [450,50,50,450]/1000.0
% endif else begin
%   Prior = [900,80,19,1]/1000.0
% endelse
\begin{tabular}{lcccc}
\hline
Character  & $\pr(H=0)$ & $\pr(H=1)$ & $\pr(H=2)$ & $\pr(H=3)$ \\  
\hline\hline
Realistic  & 0.900 & 0.080 & 0.019 & 0.001 \\
Optimistic & 0.050 & 0.100 & 0.250 & 0.600 \\
Naive      & 0.250 & 0.250 & 0.250 & 0.250 \\
\hline
\end{tabular}
\end{table}

%%%%%%%%%%%%%%%%%%%%%%%%%%%%%%%%%%%%%

%%%%%%%%%%%%%%%%%%%%%%%%%%%%%%%%%%%%%
%%%%%%%%%%%%%%%%%%%%%%%%%%%%%%%%%%%%%

If an unrealistic prior is employed, a 100\% pure sample 
of discovered lenses
can be generated by a final human classification of a subset of
robot-selected candidates. 
The cost of the optimistic robot is that in each square
degree some 670 robot class-3 candidates must be visually inspected by
a human to recover the 9 lenses present. To be 100\% complete in human
class $H=3$ systems, the robot class $\Hr = 2$ systems must be
inspected as well -- and there are some 4500 of
these. 
As well as the percentage
purities, we also show (in parentheses) on the purity charts in the figures
the predicted numbers of objects
in each bin for a 1-square degree survey area (assumed to contain
$10^4$ BRGs).
Table~\ref{tab:strategy} shows the overall completeness and
purity for various search strategies -- 
\eg\ having humans inspect all
objects with robot class $\Hr$ above some threshold --
making clear the
trade-offs involved.  We illustrate the numbers with two fiducial
imaging surveys, representing approximately what is possible now using
\hst archive images, and in the future with a space-based optical
survey telescope such as SNAP.

%%%%%%%%%%%%%%%%%%%%%%%%%%%%%%%%%%%%%%
\begin{table*}[ht!]
\caption{Lens search strategies, yields and statistics.} 
%The fiducial surveys
%and robot personalities are discussed in more detail in the text.
\label{tab:strategy}
\scriptsize
\begin{tabular*}{\linewidth}{@{\extracolsep{\fill}}lc c ccc c ccc c ccc} 
\hline\hline 
\multicolumn{2}{c}{Strategy}                               & \multicolumn{1}{c}{~} & 
\multicolumn{3}{c}{HST  yield (1 deg$^2$, $10^4$ LRGs)}    & \multicolumn{1}{c}{~} & 
\multicolumn{3}{c}{SNAP yield (1000 deg$^2$, $10^7$ LRGs)} & \multicolumn{1}{c}{~} & 
\multicolumn{3}{c}{Statistics} \\ 
%%%%%%% 
\cline{1-2} 
\cline{4-6} 
\cline{8-10} 
\cline{12-14} 
%%%%%%% 
\multicolumn{1}{l}{Character} & 
\multicolumn{1}{c}{$H_{\rm r}$ cut} & 
\multicolumn{1}{c}{} & 
\multicolumn{1}{c}{$N_{\rm cand}$} & 
\multicolumn{1}{c}{$\Delta t$} & 
\multicolumn{1}{c}{$N_{\rm lens}$} & 
\multicolumn{1}{c}{} & 
\multicolumn{1}{c}{$N_{\rm cand}$} & 
\multicolumn{1}{c}{$\Delta t$} & 
\multicolumn{1}{c}{$N_{\rm lens}$} & 
\multicolumn{1}{c}{} & 
\multicolumn{1}{c}{Rejection rate} & 
\multicolumn{1}{c}{Purity} & 
\multicolumn{1}{c}{Completeness} \\ 
%%%%%%% 
\multicolumn{1}{l}{} & 
\multicolumn{1}{c}{} & 
\multicolumn{1}{c}{} & 
\multicolumn{1}{c}{} & 
\multicolumn{1}{c}{(man-hours)} & 
\multicolumn{1}{c}{} & 
\multicolumn{1}{c}{} & 
\multicolumn{1}{c}{} & 
\multicolumn{1}{c}{(team-weeks)} & 
\multicolumn{1}{c}{} & 
\multicolumn{1}{c}{} & 
\multicolumn{1}{c}{(\%)} & 
\multicolumn{1}{c}{(\%)} & 
\multicolumn{1}{c}{(\%)} \\ 
%%%%%%% 
\cline{1-2} 
\cline{4-6} 
\cline{8-10} 
\cline{12-14} 
%%%%%%% 

realistic & $ H_{\rm r} \geq 0.5 $ &   &
399 & 1 & 5.4 &   &
399000 & 2.8 & 5400 &   &
96 & 1.4 & 54 \\
           & $ H_{\rm r} \geq 1.5 $ &   &
43 & 0 & 3.2 &   &
43000 & 0.3 & 3200 &   &
100 & 7.5 & 32 \\
           & $ H_{\rm r} \geq 2.5 $ &   &
2 & 0 & 1.9 &   &
2000 & 0 & 1900 &   &
100 & 100 & 19 \\
%%%%%%% 
\cline{1-2} 
\cline{4-6} 
\cline{8-10} 
\cline{12-14} 
%%%%%%% 
optimistic & $ H_{\rm r} \geq 0.5 $ &   &
6832 & 19 & 10 &   &
6832000 & 47.4 & 10000 &   &
32 & 0.1 & 100 \\
           & $ H_{\rm r} \geq 1.5 $ &   &
4497 & 12 & 10 &   &
4497000 & 31.2 & 10000 &   &
55 & 0.2 & 100 \\
           & $ H_{\rm r} \geq 2.5 $ &   &
672 & 2 & 8.9 &   &
672000 & 4.7 & 8900 &   &
93 & 1.3 & 89 \\
%%%%%%% 
\cline{1-2} 
\cline{4-6} 
\cline{8-10} 
\cline{12-14} 
%%%%%%% 
naive & $ H_{\rm r} \geq 0.5 $ &   &
5769 & 16 & 10 &   &
5769000 & 40.1 & 10000 &   &
42 & 0.2 & 100 \\
           & $ H_{\rm r} \geq 1.5 $ &   &
679 & 2 & 8.4 &   &
679000 & 4.7 & 8400 &   &
93 & 1.2 & 84 \\
           & $ H_{\rm r} \geq 2.5 $ &   &
89 & 0 & 6.5 &   &
89000 & 0.6 & 6500 &   &
99 & 7.3 & 65 \\
%%%%%%% 
\cline{1-2} 
\cline{4-6} 
\cline{8-10} 
\cline{12-14} 
%%%%%%% 
%%%%%%% 
\hline\hline 
\end{tabular*}
\normalsize
\vspace{\baselineskip}
\end{table*}

%%%%%%%%%%%%%%%%%%%%%%%%%%%%%%%%%%%%%

A crucial practical aspect of future large-scale surveys is the need
to cope with the extreme numbers of BRGs involved. To quantify this we
compute the rejection rate (the percentage of BRGs that are rejected
by the robot and not passed to the human classifier), and the number
of candidates needed 
to be inspected by humans. From this we estimate the
time required to carry out the human classification step.

Table~\ref{tab:strategy} shows the rejection rates and expected
classification time $\Delta t$ for our two fiducial imaging surveys,
assuming an average of 10 seconds inspection time per object, and a
team -- for the SNAP survey -- of 10 human inspectors.  We see that a
20\% complete SNAP sample generated by the realistic robot would
contain 2000 lenses and no false positives, and require negligible
classification time ($\sim 1$ hour). A $\sim90$\%-complete sample
consisting of 670,000 candidates could be generated by the optimistic
robot, and human-classified in 5 weeks. We note that the CPU time (in
2008) for
the current (and unoptimized) robot, on a 100-node compute farm, is
approximately 2 hours per square degree, or 10 weeks for the SNAP
survey.

To put this in context, the M07 search was carried out by one of us
(LAM) inspecting ``3-color'' JPG images of all 63 ACS frames, a
procedure that might be expected to be inefficient and prone to
error. Indeed, the search took 10 minutes per frame, or 60 hours per
sq degree.  At the same rate the 1000-square degree survey would take
10 full-time workers 3 years to complete.  Just targeting massive
galaxies and inspecting sequences of small cutout images leads to a
significant increase in efficiency~\citep[\eg][]{Fas++04,Fau++08}; 
however, visual inspection of every
elliptical galaxy would still take 70 weeks with the same inspection
team.  While this is a factor of 2 improvement over the EGS search
rate, the robot brings the cost down further by reducing the number of
systems needed to be human-inspected: at $\sim90$\% completeness the
BRG rejection rate is $\sim90$\%. Furthermore, we found that
displaying the candidates to the human classifier via a
``one-click'' web classification tool led to a significant reduction
in the time needed to assess each one.  Optimizing the
information at the inspectors' disposal should allow human
classification to be performed at an average rate of just a few
seconds per candidate. The survey classification time estimates 
in Table~\ref{tab:strategy}
are therefore quite conservative.

%-------------------------------------------------------------------------------

\subsection{The accuracy of the robot outputs}\label{sect:training:calib}

We now investigate the robot's modeling results for the 54 simulated
lenses classifed as $H=2$ (17 objects) and $H=3$ (37 objects). 
How accurate, and hence useful, are the lens model parameters inferred
by the robot? Since the lens model used by the robot is simpler than
that used to generate the simulations, we restrict ourselves to an
investigation of the Einstein radius, which we might hope to infer
reasonably accurately given the relatively small shears and
ellipticities involved in the simulations. We have the true,
underlying values of $\thetaE$ for each system, which we denote 
as~$\truethetaE$.

%%%%%%%%%%%%%%%%%%%%%%%%%%%%%%%%%%%%%
\begin{figure*}[!h]
\begin{minipage}{0.48\linewidth}
\centering\epsfig{file=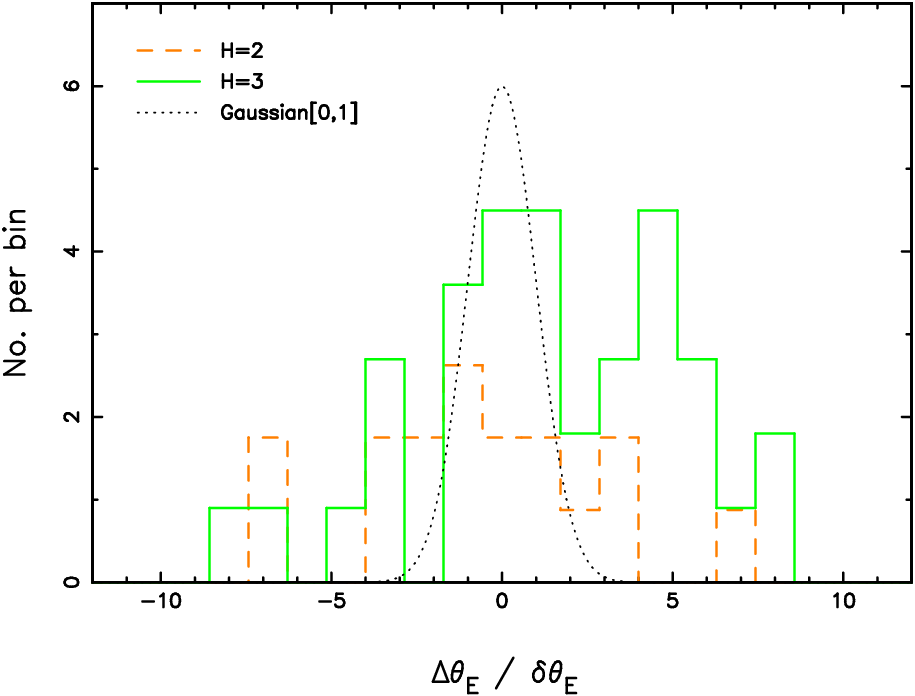,width=1.0\linewidth}
\end{minipage}\hfill
\begin{minipage}{0.48\linewidth}
\centering\epsfig{file=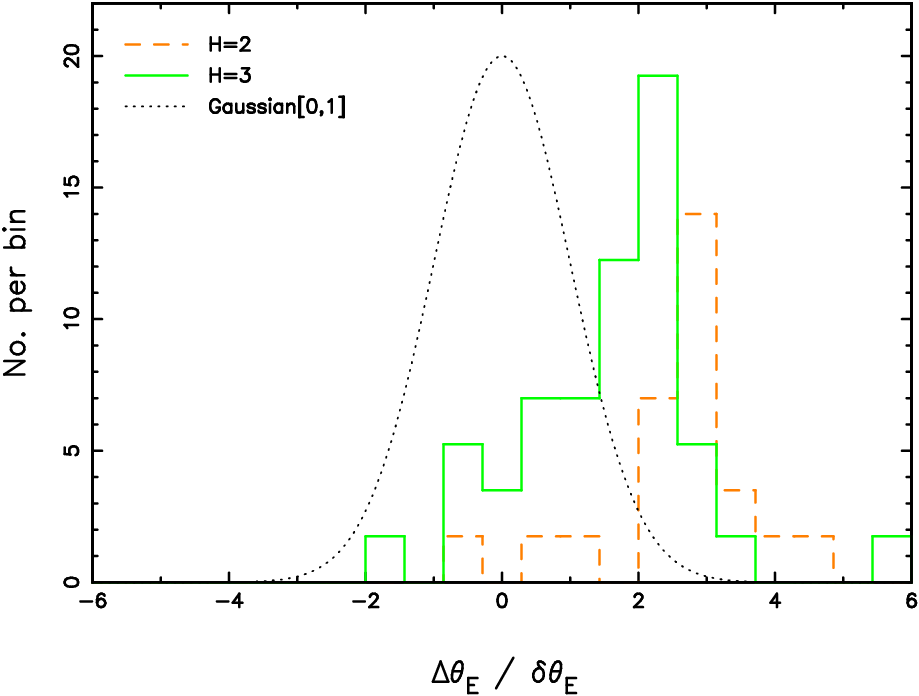,width=1.0\linewidth}
\end{minipage}
\caption{%
Left: accuracy of the model parameter~$\thetaE$ and its
``uncertainty''~$\dthetaE$.  Plotted are histograms of $\Delta
\theta_{\rm E} = (\thetaE -
\widehat{\thetaE})/\dthetaE$ for the top two human classes, compared with
a Gaussian of mean zero and width 1. 
Right: similar plot for the inferred (unlensed) source magnitude (in
the F606W band)~$\ms$ -- here there is no uncertainty $\delta \ms$,
but the unit width Gaussian still provides a useful scale for
comparison. 
\label{fig:accuracy}}
\end{figure*}
%%%%%%%%%%%%%%%%%%%%%%%%%%%%%%%%%%%%%

In the left-hand panel of Figure~\ref{fig:accuracy} we plot 
histograms
of $\Delta
\theta_{\rm E} = (\thetaE - \truethetaE)/\dthetaE$ 
for systems with human class
$H=3$ and $H=2$ (which should each look like a Gaussian centered on
zero having width unity). A peak of roughly correct width and centroid
can be seen for the robust ($H=3$) lenses, indicating that the robot's
modeling is quite meaningful in some cases but somewhat inaccurate in
others. However, the histogram has significant tails, especially at
large inferred~$\thetaE$. Some of this positive tail will be due to
the presence of external convergence in the simulated image, but only
at the few percent level. For the less convincing candidates, the
histogram is broader with a less pronounced central peak, as expected.

We might also hope to use the robot output to learn about the source
galaxy: in the right-hand panel of Figure~\ref{fig:accuracy} we plot a
similar histogram of $\Delta\ms = (\ms - \truems)$. 
While we do not infer an
uncertainty $\dms$, the unit width Gaussian still provides a useful
reference.  A $\sim$\,2~mag offset in the unlensed source magnitude
can be seen, reflecting the inability of the lens model to account for
all the flux. This is a consequence of using necessarily (to save CPU
time) inaccurate lens models: when the model does not quite predict
the image correctly, there is some mismatch between the different
multiply-imaged pixels' values. The minimum-filtering process then
leads to an underestimate of the corresponding source plane pixel
value. At the edges of the detected image features, this can lead to
an unwanted zero value in the source plane, and so to an
inferred source that is not only too faint, but also too small. The
total flux of the source is then underestimated, and the rescaling
performed before calculating $\nsigma$ is not enough to recover the
lost flux.

The robot's source magnitude estimates are therefore biased low. 
While not useful for source studies, $\ms$ is still a valid indicator
of the model quality; indeed, the discussion above shows why this so.

%-------------------------------------------------------------------------------

%%%%%%%%%%%%%%%%%%%%%%%%%%%%%%%%%%%%%%
\begin{table*}[!h]
\caption{EGS search strategies, yields and statistics.\newline 
Purity $P$, completeness $C$, and rejection rate~$R$ are all given as
percentages.}  
\label{tab:egscp}
\scriptsize
\begin{tabular}{lccccccccc}
\hline
Character  & $\Hr$ cut & $N(H=3)$ & $N(H\geq2)$ & $R$ & $P(H=3)$ & $P(H\geq2)$ & $C(H=3)$ & $C(H\geq2)$ \\ 
\hline\hline
realistic & 
$ \Hr \geq 2.5 $ & 
 0 & 
 0 & 
 100 & 
 0.0 & 
 0.0 & 
 0 & 
 0 \\
  & 
$ \Hr \geq 1.5 $ & 
 0 & 
 1 & 
 96 & 
 0.0 & 
 2.4 & 
 0 & 
 10 \\
\hline
optimistic & 
$ \Hr \geq 2.5 $ & 
 1 & 
 4 & 
 89 & 
 1.0 & 
 3.8 & 
 33 & 
 40 \\
  & 
$ \Hr \geq 1.5 $ & 
 3 & 
 9 & 
 46 & 
 0.6 & 
 1.7 & 
 100 & 
 90 \\
\hline
\hline
\end{tabular}
\normalsize
\end{table*}

%%%%%%%%%%%%%%%%%%%%%%%%%%%%%%%%%%%%%

\section{Blind testing on the EGS survey}\label{sect:egs}

Having calibrated the lens-finding robot on the training set, we now
present its application to the 60 remaining EGS ``survey''
fields~(Section~\ref{sect:training:egs}). Table~\ref{tab:egscp}
summarizes the results of this run.  All 988 lens candidates in these
fields were visually inspected by a subset of the authors (PJM, DWH,
LAM, MB, CDF), and their $H$-values simply averaged. (As noted before,
this gives non-integer values of~$H$.) The color images,
before and after lens galaxy subtraction, were displayed to aid them
in their decision-making via the cgi web form.

We ran both realistic and optimistic robots while performing the
automated classification; Table~\ref{tab:egscp} shows the statistics
from these runs.  Allowing for the small numbers involved, we find
that the completeness and purity achieved by the robot are consistent
with the predictions from the training set. With the realistic prior
we find that none of the three $H=3$ systems are recovered, consistent
with the expected low completeness (Table~\ref{tab:strategy}). With
the optimistic prior, we achieve 100\% completeness in $H=3$ systems
when setting the robot classification threshold to $\Hr \geq 1.5$, and
33\% completeness with $\Hr \geq 2.5$. That the latter is slightly
lower than expected is a reflection on the complexity of two of the
systems: as noted in M07, HST\,J141820.84+523611.2 (the ``Dewdrop'') has
a very extended source, providing a lot of confusing structure and so
fairly high values for $\Rd$ and $\dthetaE$.  HST\,J141735.73+522646.3
(the ``Cross'') has almost point-like images in an asymmetric pattern,
caused by the combination of both internal ellipticity and external
shear in the lens~(\eg\ M07).  This leads to a poor model fit and a
high value for $\nsigma$. The rejection ratios ensuing from the
optimistic search strategies (46 and 89\%) match well the predictions
from the training set in Table~\ref{tab:strategy} (55 and 93\%).

%%%%%%%%%%%%%%%%%%%%%%%%%%%%%%%%%%%%%
\begin{figure*}[!ht]
\centering\epsfig{file=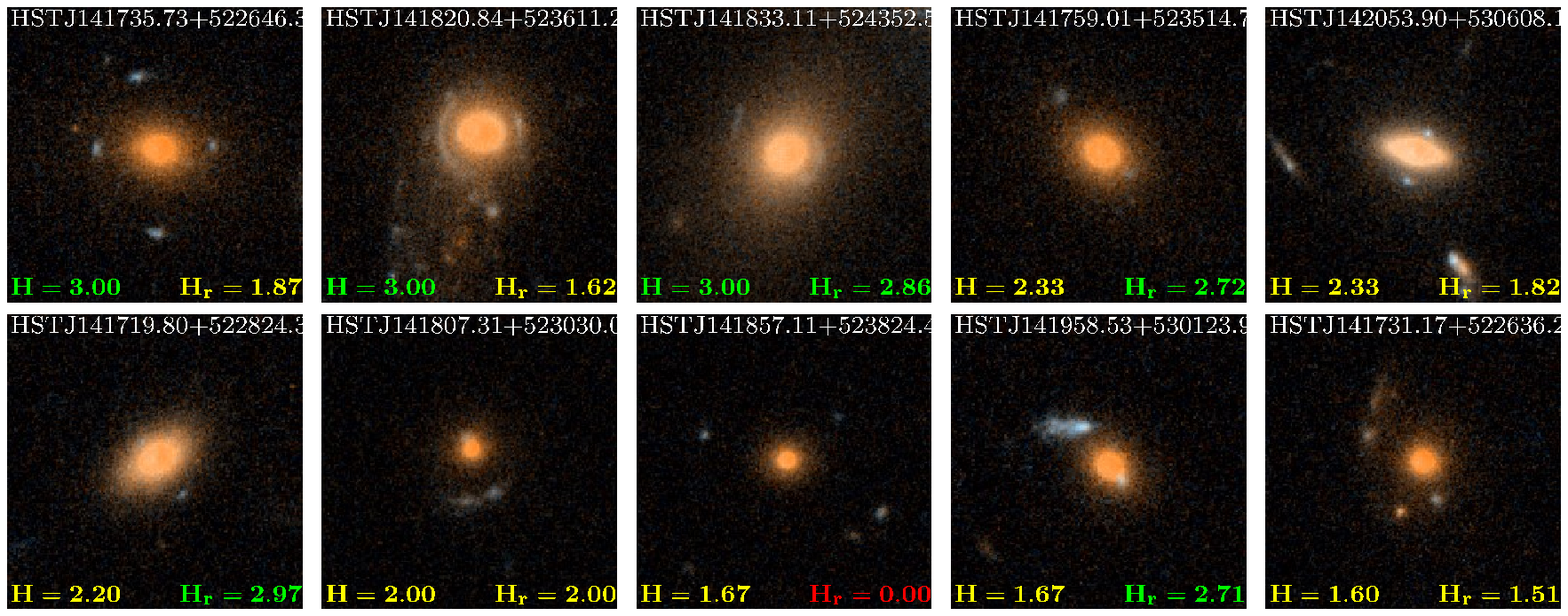,width=1.0\linewidth}
\caption{%
Automated lens search candidates in the EGS survey area.  Candidates
are sorted (left to right, top to bottom) by their human
classification parameter,~$H$ -- all 10 objects with $H > 1.5$ are
shown.  Both human and robot classification parameter ($\Hr$) are
shown overlaid on the images, with color scheme green $\rightarrow
H=3$, amber $\rightarrow H=2$ and red $\rightarrow H=1$. All cutout
images are 6\,arcsec on a side.
\label{fig:candidates}}
\end{figure*}
%%%%%%%%%%%%%%%%%%%%%%%%%%%%%%%%%%%%%

%%%%%%%%%%%%%%%%%%%%%%%%%%%%%%%%%%%%%
\begin{table*}[!ht]
\caption{Robot outputs for the 10 lens candidates in
  Figure~\ref{fig:candidates}.} 
\label{tab:candidates}
\scriptsize
\begin{tabular}{cccccccccc}
\hline
Object name  & Position in                    & 
  $\nsigma$  & $\thetaE$  &  $\dthetaE$ & 
  $\Rd$  &  Source magnitude  & Robot class & Human class \\  
             & Figure~\ref{fig:candidates} & 
             & (arcsec)   &  (arcsec)   & 
         &  AB, F606W  & $(\Hr)$ & $(H)$ \\ 
\hline\hline
HSTJ141735.73+522646.3 & (1,1) & 13.23 & 0.96 & 0.07 & 0.60 & 30.42 & 1.87 & 3.00 \\
HSTJ141820.84+523611.2 & (2,1) & 4.08 & 0.66 & 0.14 & 0.69 & 27.92 & 1.62 & 3.00 \\
HSTJ141833.11+524352.5 & (3,1) & 4.11 & 0.84 & 0.06 & 0.62 & 28.40 & 2.86 & 3.00 \\
\hline
HSTJ141759.01+523514.7 & (4,1) & 5.07 & 0.96 & 0.06 & 0.57 & 28.84 & 2.72 & 2.33 \\
HSTJ142053.90+530608.1 & (5,1) & 7.19 & 0.54 & 0.04 & 0.31 & 29.56 & 1.82 & 2.33 \\
HSTJ141719.80+522824.3 & (1,2) & -1.18 & 0.60 & 0.04 & 0.75 & 29.62 & 2.97 & 2.20 \\
HSTJ141807.31+523030.0 & (2,2) & 1.69 & 0.66 & 0.16 & 1.00 & 29.29 & 2.00 & 2.00 \\
HSTJ141857.11+523824.4 & (3,2) & 99.00 & 0.00 & 99.00 & 0.00 & 99.00 & 0.00 & 1.67 \\
HSTJ141958.53+530123.9 & (4,2) & 1.21 & 0.66 & 0.07 & 0.48 & 27.96 & 2.71 & 1.67 \\
HSTJ141731.17+522636.2 & (5,2) & 21.31 & 1.08 & 0.07 & 0.59 & 28.75 & 1.51 & 1.60 \\
\hline
\end{tabular}
\end{table*}

%%%%%%%%%%%%%%%%%%%%%%%%%%%%%%%%%%%%%

Figure~\ref{fig:candidates} shows all 10 objects with human class
$H \geq 2.5$ (the ``A-list'', three objects) or $1.5 < H < 2.5$ 
(the ``B-list'', seven
objects). The robot output data are tabulated in
Table~\ref{tab:candidates}. 
The B-list systems include the three that were noted during the M06
\& M07 by-eye searches -- four are new lens candidates (albeit of low
quality).  
One B-list system was rejected by the robot, a result of
its large image separation.

These four new B-list candidates illustrate an important point,
namely that automated searches ameliorate the problem of human error
in a by-eye search. By focusing on the few galaxies consistent with
being lenses, the distractions of the rest of the image are removed
and there is less chance of missing an interesting object.

%-------------------------------------------------------------------------------

\section{Discussion}\label{sect:discuss}

In this Section we first identify several areas where our robot's
performance could be improved, and then compare our approach with
others suggested in the literature.

% - - - - - - - - - - - - - - - - - - - - - - - - - - - - - - - - - - - - - -

\subsection{A more extensive training dataset}

The blind test on real survey data described in Section~\ref{sect:egs}
suggests that our simulations are sufficiently realistic to give us an
accurate PDF for lens candidate classification, although a larger
sample of known lenses would assist here. For example, the \hst/ACS
images taken during the SLACS survey would provide a training set of
some 70 galaxy-galaxy strong lenses \citep{Bol++08}. However, we must
be careful to match the robot's training set with the kinds of lenses
expected to be {\it found in high resolution imaging data}. The
selection function of the SLACS lenses is quite different, favoring
low redshift lens galaxies and high magnification image configurations
(Einstein rings).  While it may be argued that the latter is a
desirable property of the target lenses, if we seek to find all lenses
we need to educate the robot accordingly. One option for future work
would be to use the entire EGS survey (as well as the simulated
lenses) as a training set -- although some care may be needed when
applying the robot 
to new data of different depth and resolution.  If our
training set is assumed adequate, then two sources of incompleteness
and inefficiency remain. One is the treatment of the training set, and
the other is the lens modeling itself.

% - - - - - - - - - - - - - - - - - - - - - - - - - - - - - - - - - - - - - -

\subsection{More accurate PDF modeling}

It can be seen in Figures~\ref{fig:pdf0}--\ref{fig:pdf3} 
that there are a few outliers to the derived
PDFs, indicating that these model
distributions are a somewhat lossy compression of the
information in the training set. One way of correcting this would be
to increase the complexity, and therefore inclusiveness, of the PDF
models to reduce the number of outliers. However, the problem of
degeneracy between the 4 different PDFs remains -- this can be broken
by increasing the dimensionality of the robot output data-space. While
an exhaustive investigation of the individual cases is beyond the
scope of this paper, we make the following general suggestion for
future work. Most of the bright red objects passed to the robot for
lens-modeling are {\it not} massive elliptical galaxies; these should
give complex residual images even when the lensing-consistent flux is
subtracted. Some quantitative morphology analysis of such maps might
produce a useful extra dimension to use in ruling out
non-lenses. Likewise, the pre-selection of BRG candidates could be
improved, using some measure of concentration to favor the massive
galaxies.

One disadvantage of the approach taken here is that we treat the four
classifications ($H=0,1,2,3$) as exclusive and unrelated ``bins'' into
which lens candidates fall; that is, we make no use of the fact that
there is really a continuum from $H=0$ to $H=3$, and that $H=1$ is
between $H=0$ and $H=2$, and that $H=0$ is very far from $H=3$.  A
more sophisticated approach would make use of this continuity
information, perhaps by working with the five-dimensional distribution
of the four scalars \emph{and}~$H$; that is, treating~$H$ as a fifth
scalar to be predicted using the other four.  A larger training set,
or a training set classified by a larger number of humans, could also
effectively increase the granularity of the $H$-statistic and provide
a better basis for treating the classification as a continuum rather
than a discontinuous set of exclusive and unrelated bins.

% - - - - - - - - - - - - - - - - - - - - - - - - - - - - - - - - - - - - - -

\subsection{More accurate lens modeling}

The lens modeling itself could be made cleaner at the same time as
improving this pre-selection: fitting the BRG light with a bulge+disk
light profile may better
suppress the symmetrical disk-like residuals that
can mimic lensed images, while providing some quantitative estimate of
galaxy type (and hence mass). In the future, with surveys in many
filters we can hope to extend this modeling to include the photometric
redshift and stellar population properties of the BRG, and use the
fundamental plane to estimate the BRG mass directly.

While the above desiderata may improve the efficiency of the lens
search, they all favor smooth, clean lens galaxies, and sparsely
populated source planes.  This approach is somewhat justified for some
of our strong lensing science goals, \ie\ those that require samples
of massive regular elliptical galaxies whose lensing effects are
easily modeled. In fact, one of the attractive aspects of the future
imaging surveys is that they will be capable of sampling further down
the lensing cross-section distribution, to where the lens galaxies are
higher redshift, and/or less bulge-dominated. We have shown that our
approach can deal with massive galaxies that do have, for example,
disk components, but at the cost of enlarging the size of the visual
inspection candidate list. If we want to find more exotic disky or
complex lenses, then the robot's modeling capabilities must be
increased accordingly. The pre-selection can likely be made much more
stringent (\eg\ selecting close pairs of BRGs) -- in which case a
purely visual inspection may be the most effective strategy. However,
we already see in the case of the EGS lenses that more than three lens
model parameters are required for a good fit to the data: more work is
required on making such flexible lens fits feasible in the available
CPU time.

An unwelcome side-effect of poor lens-modeling is that the output
parameters may not be useful for some other desirable follow-on
work. For example, the source plane photometry performed in this work
is not accurate enough to estimate the photometric redshift of the
source. This should probably not be a concern -- even $10^4$ new
lenses, once found in the SNAP survey for example, could be re-modeled
straightforwardly given the computing power assumed in
Table~\ref{tab:strategy}.

Finally, it is worth revisiting the assumptions made in our lens
modeling algorithm. Perhaps the most important is our neglect of the
PSF. For the high resolution image surveys we have restricted
ourselves to, we expect this not to be a problem for the more numerous
extended galaxy-source lenses.  However, we should not be surprised to
find the robot failing to detect lensed quasars, especially if they
are very bright. A more advanced lens modeler would have to
incorporate some form of deconvolution; the most stable way to do this
is to work with a model source and infer its parameters by predicting
the image plane. However, this will add parameters to the model and
prevent us using the efficient ``minimum-filtering''
scheme. Nevertheless, detecting bright lensed quasars would require
the PSF to be taken into account properly, as would extending our
approach to ground-based data, with its larger PSF width to source
size ratio.

% - - - - - - - - - - - - - - - - - - - - - - - - - - - - - - - - - - - - - -

\subsection{A more objective classification scheme}

In order to bypass the more straightforwardly inferred human
classification parameter, and instead answer the question 
``Is this a
gravitational lens or not?'' we would have to do the following: for
each conceivably massive, distant galaxy in a large imaging survey,
search the entire space of all reasonable models for that galaxy's
lensing potential, and search the entire space of reasonable
distributions of resolved and unresolved background sources in angle,
redshift, and color, looking for a model to explain the morphology
observed in and around the galaxy image.  If, in this enormous space,
there is a reasonable model for the potential and a reasonable model
for the background sources that explains a significant and not
fine-tuned portion of the intensity in and around the image of the
lensing galaxy, then that galaxy is a very strong candidate for a
multiply-imaging lens. The null model against which this would be
compared would be that where all features in the elliptical-subtracted
image are assumed unlensed. The key point is that the lensing
hypothesis is potentially simpler for comparable (and perhaps better)
goodness of fit, since fewer individual sources need to be fitted.

In practice, it is not currently possible to implement this scheme in
full. As previously discussed, at present it is not
practical to perform fully general lens modeling for every
object. Perhaps more importantly, there is also currently no reliable
way to determine which parts of the image of a putative lensing galaxy
are part of that galaxy (or foreground) and which originate from
background sources. For this we would need a comprehensive
understanding of galaxy morphology quantified as a complex joint prior
for the morphology parameters. While there are promising signs of such
a formalism being developed \citep[\eg][]{LPM04,Mas++04a}, we are not
there yet. This means that, at present, there is no quantitative
meaning to the words ``reasonable,'' ``significant,'' and ``not
fine-tuned,'' and so the usual evidence ratio used for model selection
is not available to us.

Nonetheless, the method we have described here represents a first
attempt at the model-oriented lens searching scheme, including a
number of necessary approximations and simplifications. Its extension
to more powerful models and fitting algorithms is straightforward, and
we leave this to future work. The quantification of non-lens galaxy
morphology presents a greater challenge.

% - - - - - - - - - - - - - - - - - - - - - - - - - - - - - - - - - - - - - -

\subsection{Comparison with other automated lens detection algorithms}

Most of the automated lens detection schemes proposed to date have
focused on finding curved arc-like structures
\citep{LSS05,S+B07,Ala07,K+D08}. These have the benefit of finding
lenses by their sources, not their lens galaxies -- an important
distinction for lens-statistics studies \citep[\eg][]{Kee02,Koc06} and
for finding dark gravitational lenses. While none of these methods
rely on the subtraction of the lens galaxy light before applying the
algorithms, it would seem profitable to do so when searching for
galaxy-scale lenses. All include a final human inspection step to
provide quality control.

The Ringfinder algorithm~\citep[Gavazzi et al in
preparation,][]{Cab++07} does include the subtraction of the lens
light: as this is modeled by rescaling the reddest available image,
the method is restricted to multi-filter data.  It was designed for
the CFHT legacy survey ground-based data, and so represents the first
attempt at specifically finding galaxy-scale lenses robotically, from
the ground. While the analysis of the blue residuals is more ad-hoc,
the results could in principle be trained in the same way we have
described here.

The arc-detection algorithm used by \citet{Est++07} in searching the
SDSS images shares a key feature with our robot: they use a neural
network to assess their (different) 4 data that describe each
candidate arc. The probabilistic framework presented here can itself
be viewed as a simple machine-learning process.  Although crude (and
in fact, hand-made), our framework does have the important benefit of
providing some the insight into the problem. A fruitful line of future
enquiry could be to replace our robot's PDF (with its dependence on a
limited data vector~$\data$) with a neural network; the natural
extension of this would then be to increase the number of inputs to be
the pixels of the image itself.  This approach would perhaps solve the
problem of the non-lens galaxy morphology as well.

%-------------------------------------------------------------------------------

\section{Conclusions}\label{sect:concl}

We have developed a novel approach to the automated detection and
classification of strong galaxy-scale gravitational lenses in high
resolution imaging surveys. After training our software robot on
simulated and real \hst data we draw the following conclusions:

\begin{itemize}

\item For high resolution data and sufficiently faint and extended
  images, we can neglect the PSF and reduce the complexity of the lens
  model to three parameters, generating the unlensed source plane by a
  simple and robust ray-tracing and minimum-filtering scheme. In this
  way, our
  robot is able to return crude lens models that predict the
  images seen in both simulated and real galaxy-scale lenses.

\item While the Einstein radius and source magnitude returned by the
  robot are not yet accurate enough for further use, improvements in
  the lens modeling should make these parameters useful. The automated
  nature of the detection process means that the selection function is
  well-defined, such that measuring \eg\ $dn/d\thetaE$ should be
  meaningful.

\item Using a set of data derived from the lens model, we infer for
  each candidate the classification parameter~$H$ that a human
  inspector would have assigned it. This is a well-defined procedure
  that can be calibrated straightforwardly 
  using a large sample of simulated lenses and
  known non-lenses; the calibration information is compressed as a set
  of PDFs whose estimation comprises the ``training'' of the
  robot. While some systems in the training set remain as outliers to
  the model distribution, this does not have a catastrophic effect on
  the automated classification.

\item The completeness and purity of any survey are partly determined
  by the prior PDF on the classification parameter, in our case
  $\pr(H)$. A realistic prior distribution of $H$-values heavily
  favors the classification $H=0$ (on the grounds that they are known
  to be much more common), and predicts higher class objects to be
  correspondingly rare: it makes the search efficient, with any
  loss of completeness being
  due to the inadequacies of the modeling
  process.  Setting a threshold of $\Hr \geq 2.5$, we find a purity of
  $\sim100$\% at a completeness of $\sim20$\%.

\item We can choose to improve the completeness at
  the expense of the purity and efficiency by changing the prior PDF,
  which is the analog of employing a classifier of different
  {\it character}.  A more ``optimistic'' classifier would be happy
  to see many more high class objects, and at least at present 
  this seems
  necessary to achieve 100\% completeness. The price of high
  completeness is a low rejection rate: with the optimistic robot
  threshold set to $\Hr \geq 2.5$, we find a rejection rate of
  $\sim90$\% at a completeness of $\sim90$\%.
 
\item A realistic robot may be most appropriate for future large
  imaging surveys where human inspection is costly. A 1000 square
  degree survey with a space telescope such as SNAP would contain
  $\sim10^7$ BRGs and $\sim10^4$ lenses; the current 
  realistic robot's sample
  would comprise 2000 of these, with no inspection required.
 
\item An optimistic robot is more appropriate for present-day lens
  searches, where the numbers are small enough for human inspection to
  be cheap, and where every new lens counts.  A search area
  of 1 square
  degree, such as that provided by the \hst/ACS archive, would lead to
  an optimistic robot sample of $\sim5000$ candidates, with the human
  classification taking $\sim1$~day.

\item Applying the optimistic robot to the EGS survey, we recovered
  all three human class-3 lenses, and all but one of the three
  human class-2 lens
  candidates from M06 and M07. We also discovered four new human
  class-2 lens candidates.

\end{itemize}

At the time of writing, the era of wide-field imaging from space is
still $\sim1$ decade away. Continuing to train our software robot on
\hst data, we should be optimistic about the prospects of a feasible
search generating a well-defined statistical sample of galaxy-scale
strong lenses from an imaging survey like that of SNAP. The approach
we have described here is maximally informative, in that it
incorporates our expectations about the typical lenses in the
universe. However, perhaps the 
biggest challenge will be discovering the more
complex and unexpected strong lenses yet to be seen.

%-------------------------------------------------------------------------------

\acknowledgments

We thank Raphael Gavazzi, Tommaso Treu, Eric Morganson, 
Elisabeth Newton, Marco
Lombardi, Ole Moller and Konrad Kuijken for useful discussions,
Jean-Paul Kneib for much encouragement, and
Cecile Faure for a careful reading of the manuscript. We are grateful
to the EGS team for providing their high level science products at an
early stage in the project.  DWH thanks the staff of the Spitzer
Science Center for their hospitality during his visit when some of
this work was carried out.  
Support for this work was provided by NASA through
grant number HST-AR-11289 (the HAGGLeS project) from the Space Telescope
Science Institute, which is operated by AURA, Inc., under NASA contract NAS 5-26555.
The work of PJM, RDB and MB was 
supported in part by
the U.S.\ Department of Energy, under contract number DE-AC02-76SF00515 
at the Stanford Linear Accelerator Center. The work of RDB was supported in part by a
National Science Foundation grant, ``Gravitational optics, 
dark matter, and the evolution of faint galaxies,'' and by a 
U.S.\ Department of Energy Computational Astrophysics Consortium grant, 
``3. supernovae, gamma-ray bursts, and nucleosynthesis".
The work of PJM was supported 
by the TABASGO foundation in the form of a research
fellowship. The work of LAM was carried out at the Jet Propulsion
Laboratory, California Institute of Technology, under a contract with
NASA.

%-------------------------------------------------------------------------------

\bibliographystyle{apj}

%-------------------------------------------------------------------------------

\end{document}